\RequirePackage{snapshot}
\documentclass[11pt,twoside,british]{article}
\usepackage[T1]{fontenc}
\usepackage[latin9]{inputenc}
\usepackage[a4paper]{geometry}
\usepackage{fancyhdr}
\usepackage{color}
\usepackage{babel}
\usepackage{verbatim}
\usepackage{refstyle}
\usepackage{booktabs}
\usepackage{mathtools}
\usepackage{enumitem}
\usepackage{amsmath}
\usepackage{amsthm}
\usepackage{amssymb}
\usepackage{setspace}
\usepackage[authoryear,longnamesfirst]{natbib}
\usepackage{xargs}[2008/03/08]
\usepackage[bookmarks=true,bookmarksnumbered=false,bookmarksopen=false,
 breaklinks=false,pdfborder={0 0 0},pdfborderstyle={},backref=false,colorlinks=false]
 {hyperref}

\makeatletter

\AtBeginDocument{\providecommand\eqref[1]{\ref{eq:#1}}}
\AtBeginDocument{\providecommand\secref[1]{\ref{sec:#1}}}
\AtBeginDocument{\providecommand\assref[1]{\ref{ass:#1}}}
\AtBeginDocument{\providecommand\thmref[1]{\ref{thm:#1}}}
\AtBeginDocument{\providecommand\appref[1]{\ref{app:#1}}}
\AtBeginDocument{\providecommand\propref[1]{\ref{prop:#1}}}
\AtBeginDocument{\providecommand\lemref[1]{\ref{lem:#1}}}
\AtBeginDocument{\providecommand\tblref[1]{\ref{tbl:#1}}}
\AtBeginDocument{\providecommand\subsecref[1]{\ref{subsec:#1}}}
\providecommand{\tabularnewline}{\\}
\RS@ifundefined{subsecref}
  {\newref{subsec}{name = \RSsectxt}}
  {}
\RS@ifundefined{thmref}
  {\def\RSthmtxt{theorem~}\newref{thm}{name = \RSthmtxt}}
  {}
\RS@ifundefined{lemref}
  {\def\RSlemtxt{lemma~}\newref{lem}{name = \RSlemtxt}}
  {}

\numberwithin{equation}{section}
\numberwithin{figure}{section}
\numberwithin{table}{section}
\theoremstyle{remark}
\newtheorem*{notation*}{\protect\notationname}
\theoremstyle{plain}
\newtheorem{assumption}{\protect\assumptionname}
\theoremstyle{plain}
\newtheorem{prop}{\protect\propositionname}[section]
\theoremstyle{plain}
\newtheorem{thm}{\protect\theoremname}[section]
\theoremstyle{plain}
\newtheorem{lem}{\protect\lemmaname}[section]
\theoremstyle{definition}
\newtheorem{example}{\protect\examplename}[section]
\theoremstyle{remark}
\newtheorem{rem}{\protect\remarkname}[section]

\setlist[enumerate,1]{label=\upshape{(\roman*)}, ref=(\roman*)}
\setlist[enumerate,2]{label=\upshape{(\alph*)}, ref=(\alph*)}
\setlist[enumerate,3]{label=\upshape{\roman*.}, ref=\roman*}

\makeatletter
  \@ifpackageloaded{refstyle}{}{\usepackage{refstyle}}
\makeatother

\newref{thm}{name = Theorem~}
\newref{prop}{name = Proposition~}
\newref{lem}{name = Lemma~}
\newref{cor}{name = Corollary~}
\newref{ass}{name = }
\newref{exa}{name = Example~}
\newref{rem}{name = Remark~}
\newref{def}{name = Definition~}

\newref{eq}{refcmd = (\ref{#1})}

\newref{sec}{name = Section~}
\newref{sub}{name = Section~}
\newref{subsec}{name = Section~}
\newref{app}{name = Appendix~}
\newref{supp}{name = Appendix~}

\newref{fig}{name = Figure~}
\newref{tbl}{name = Table~}

\makeatletter
  \@ifpackageloaded{mathrsfs}{}{\usepackage{mathrsfs}}
\makeatother

\date{}

\usepackage{nextpage}

\allowdisplaybreaks[1]

\usepackage{scalefnt}
\usepackage[group-separator={,}]{siunitx}
\newcommand\smaller[2][0.85]{{\scalefont{#1}#2}}
\newcommand{\ass}[1]{{\upshape{\smaller[0.76]{#1}}}}

\newcommand{\assumpname}[1]{%
  \renewcommand{\theassumption}{\ass{#1}}%
}

\makeatletter
\newsavebox{\@brx}
\newcommand{\dbllangle}[1][]{\savebox{\@brx}{\(\m@th{#1\langle}\)}%
  \mathopen{\copy\@brx\kern-0.5\wd\@brx\usebox{\@brx}}}
\newcommand{\dblrangle}[1][]{\savebox{\@brx}{\(\m@th{#1\rangle}\)}%
  \mathclose{\copy\@brx\kern-0.5\wd\@brx\usebox{\@brx}}}
\makeatother

\usepackage{changepage}

\theoremstyle{definition}
\newtheorem{examplex}{Example}
\renewenvironment{example}
  {\pushQED{\qed}\examplex}
  {\popQED\endexamplex}

\numberwithin{examplex}{section}

\newcounter{subremark}[rem]

\newcounter{savedexnumber}

\usepackage{mathdots}

\usepackage[usestackEOL]{stackengine}
\setstackgap{S}{5pt}
\newcommand{\authaffil}[2]{\Shortunderstack{#1\\\small{#2}}}

\usepackage{needspace}

\usepackage{bibunits}
\defaultbibliography{cksvar}
\defaultbibliographystyle{ecta}

\defcitealias{DMW22}{DMW25}

\usepackage{booktabs}

\makeatother

\providecommand{\assumptionname}{Assumption}
\providecommand{\examplename}{Example}
\providecommand{\lemmaname}{Lemma}
\providecommand{\notationname}{Notation}
\providecommand{\propositionname}{Proposition}
\providecommand{\remarkname}{Remark}
\providecommand{\theoremname}{Theorem}

\begin{document}


\global\long\def\uwrite#1#2{\underset{#2}{\underbrace{#1}} }%

\global\long\def\blw#1{\ensuremath{\underline{#1}}}%

\global\long\def\abv#1{\ensuremath{\overline{#1}}}%

\global\long\def\vect#1{\mathbf{#1}}%


\global\long\def\smlseq#1{\{#1\} }%

\global\long\def\seq#1{\left\{  #1\right\}  }%

\global\long\def\smlsetof#1#2{\{#1\mid#2\} }%

\global\long\def\setof#1#2{\left\{  #1\mid#2\right\}  }%


\global\long\def\goesto{\ensuremath{\rightarrow}}%

\global\long\def\ngoesto{\ensuremath{\nrightarrow}}%

\global\long\def\uto{\ensuremath{\uparrow}}%

\global\long\def\dto{\ensuremath{\downarrow}}%

\global\long\def\uuto{\ensuremath{\upuparrows}}%

\global\long\def\ddto{\ensuremath{\downdownarrows}}%

\global\long\def\ulrto{\ensuremath{\nearrow}}%

\global\long\def\dlrto{\ensuremath{\searrow}}%


\global\long\def\setmap{\ensuremath{\rightarrow}}%

\global\long\def\elmap{\ensuremath{\mapsto}}%

\global\long\def\compose{\ensuremath{\circ}}%

\global\long\def\cont{C}%

\global\long\def\cadlag{D}%

\global\long\def\Ellp#1{\ensuremath{\mathcal{L}^{#1}}}%


\global\long\def\naturals{\ensuremath{\mathbb{N}}}%

\global\long\def\reals{\mathbb{R}}%

\global\long\def\complex{\mathbb{C}}%

\global\long\def\rationals{\mathbb{Q}}%

\global\long\def\integers{\mathbb{Z}}%


\global\long\def\abs#1{\ensuremath{\left|#1\right|}}%

\global\long\def\smlabs#1{\ensuremath{\lvert#1\rvert}}%
 
\global\long\def\bigabs#1{\ensuremath{\bigl|#1\bigr|}}%
 
\global\long\def\Bigabs#1{\ensuremath{\Bigl|#1\Bigr|}}%
 
\global\long\def\biggabs#1{\ensuremath{\biggl|#1\biggr|}}%

\global\long\def\norm#1{\ensuremath{\left\Vert #1\right\Vert }}%

\global\long\def\smlnorm#1{\ensuremath{\lVert#1\rVert}}%
 
\global\long\def\bignorm#1{\ensuremath{\bigl\|#1\bigr\|}}%
 
\global\long\def\Bignorm#1{\ensuremath{\Bigl\|#1\Bigr\|}}%
 
\global\long\def\biggnorm#1{\ensuremath{\biggl\|#1\biggr\|}}%

\global\long\def\floor#1{\left\lfloor #1\right\rfloor }%
\global\long\def\smlfloor#1{\lfloor#1\rfloor}%

\global\long\def\ceil#1{\left\lceil #1\right\rceil }%
\global\long\def\smlceil#1{\lceil#1\rceil}%


\global\long\def\Union{\ensuremath{\bigcup}}%

\global\long\def\Intsect{\ensuremath{\bigcap}}%

\global\long\def\union{\ensuremath{\cup}}%

\global\long\def\intsect{\ensuremath{\cap}}%

\global\long\def\pset{\ensuremath{\mathcal{P}}}%

\global\long\def\clsr#1{\ensuremath{\overline{#1}}}%

\global\long\def\symd{\ensuremath{\Delta}}%

\global\long\def\intr{\operatorname{int}}%

\global\long\def\cprod{\otimes}%

\global\long\def\Cprod{\bigotimes}%


\global\long\def\smlinprd#1#2{\ensuremath{\langle#1,#2\rangle}}%

\global\long\def\inprd#1#2{\ensuremath{\left\langle #1,#2\right\rangle }}%

\global\long\def\orthog{\ensuremath{\perp}}%

\global\long\def\dirsum{\ensuremath{\oplus}}%


\global\long\def\spn{\operatorname{sp}}%

\global\long\def\rank{\operatorname{rk}}%

\global\long\def\proj{\operatorname{proj}}%

\global\long\def\tr{\operatorname{tr}}%

\global\long\def\vek{\operatorname{vec}}%

\global\long\def\diag{\operatorname{diag}}%

\global\long\def\col{\operatorname{col}}%


\global\long\def\smpl{\ensuremath{\Omega}}%

\global\long\def\elsmp{\ensuremath{\omega}}%

\global\long\def\sigf#1{\mathcal{#1}}%

\global\long\def\sigfield{\ensuremath{\mathcal{F}}}%
\global\long\def\sigfieldg{\ensuremath{\mathcal{G}}}%

\global\long\def\flt#1{\mathcal{#1}}%

\global\long\def\filt{\mathcal{F}}%
\global\long\def\filtg{\mathcal{G}}%

\global\long\def\Borel{\ensuremath{\mathcal{B}}}%

\global\long\def\cyl{\ensuremath{\mathcal{C}}}%

\global\long\def\nulls{\ensuremath{\mathcal{N}}}%

\global\long\def\gauss{\mathfrak{g}}%

\global\long\def\leb{\mathfrak{m}}%


\global\long\def\prob{\ensuremath{\mathbb{P}}}%

\global\long\def\Prob{\ensuremath{\mathbb{P}}}%

\global\long\def\Probs{\mathcal{P}}%

\global\long\def\PROBS{\mathcal{M}}%

\global\long\def\expect{\ensuremath{\mathbb{E}}}%

\global\long\def\Expect{\ensuremath{\mathbb{E}}}%

\global\long\def\probspc{\ensuremath{(\smpl,\filt,\Prob)}}%


\global\long\def\iid{\ensuremath{\textnormal{i.i.d.}}}%

\global\long\def\as{\ensuremath{\textnormal{a.s.}}}%

\global\long\def\asp{\ensuremath{\textnormal{a.s.p.}}}%

\global\long\def\io{\ensuremath{\ensuremath{\textnormal{i.o.}}}}%

\newcommand\independent{\protect\mathpalette{\protect\independenT}{\perp}}
\def\independenT#1#2{\mathrel{\rlap{$#1#2$}\mkern2mu{#1#2}}}

\global\long\def\indep{\independent}%

\global\long\def\distrib{\ensuremath{\sim}}%

\global\long\def\distiid{\ensuremath{\sim_{\iid}}}%

\global\long\def\asydist{\ensuremath{\overset{a}{\distrib}}}%

\global\long\def\inprob{\ensuremath{\overset{p}{\goesto}}}%

\global\long\def\inprobu#1{\ensuremath{\overset{#1}{\goesto}}}%

\global\long\def\inas{\ensuremath{\overset{\as}{\goesto}}}%

\global\long\def\eqas{=_{\as}}%

\global\long\def\inLp#1{\ensuremath{\overset{\Ellp{#1}}{\goesto}}}%

\global\long\def\indist{\ensuremath{\overset{d}{\goesto}}}%

\global\long\def\eqdist{=_{d}}%

\global\long\def\wkc{\ensuremath{\rightsquigarrow}}%

\global\long\def\wkcu#1{\overset{#1}{\ensuremath{\rightsquigarrow}}}%

\global\long\def\plim{\operatorname*{plim}}%


\global\long\def\var{\operatorname{var}}%

\global\long\def\lrvar{\operatorname{lrvar}}%

\global\long\def\cov{\operatorname{cov}}%

\global\long\def\corr{\operatorname{corr}}%

\global\long\def\bias{\operatorname{bias}}%

\global\long\def\MSE{\operatorname{MSE}}%

\global\long\def\med{\operatorname{med}}%

\global\long\def\avar{\operatorname{avar}}%

\global\long\def\se{\operatorname{se}}%

\global\long\def\sd{\operatorname{sd}}%


\global\long\def\nullhyp{H_{0}}%

\global\long\def\althyp{H_{1}}%

\global\long\def\ci{\mathcal{C}}%


\global\long\def\simple{\mathcal{R}}%

\global\long\def\sring{\mathcal{A}}%

\global\long\def\sproc{\mathcal{H}}%

\global\long\def\Wiener{\ensuremath{\mathbb{W}}}%

\global\long\def\sint{\bullet}%

\global\long\def\cv#1{\left\langle #1\right\rangle }%

\global\long\def\smlcv#1{\langle#1\rangle}%

\global\long\def\qv#1{\left[#1\right]}%

\global\long\def\smlqv#1{[#1]}%


\global\long\def\trans{\mathsf{T}}%

\global\long\def\indic{\ensuremath{\mathbf{1}}}%

\global\long\def\Lagr{\mathcal{L}}%

\global\long\def\grad{\nabla}%

\global\long\def\pmin{\ensuremath{\wedge}}%
\global\long\def\Pmin{\ensuremath{\bigwedge}}%

\global\long\def\pmax{\ensuremath{\vee}}%
\global\long\def\Pmax{\ensuremath{\bigvee}}%

\global\long\def\sgn{\operatorname{sgn}}%

\global\long\def\argmin{\operatorname*{argmin}}%

\global\long\def\argmax{\operatorname*{argmax}}%

\global\long\def\Rp{\operatorname{Re}}%

\global\long\def\Ip{\operatorname{Im}}%

\global\long\def\deriv{\ensuremath{\mathrm{d}}}%

\global\long\def\diffnspc{\ensuremath{\deriv}}%

\global\long\def\diff{\ensuremath{\,\deriv}}%

\global\long\def\i{\ensuremath{\mathrm{i}}}%

\global\long\def\e{\mathrm{e}}%

\global\long\def\sep{,\ }%

\global\long\def\defeq{\coloneqq}%

\global\long\def\eqdef{\eqqcolon}%

\global\long\def\err{\varepsilon}%

\global\long\def\mset#1{\mathcal{#1}}%

\global\long\def\largedec#1{\mathbf{#1}}%

\global\long\def\z{\largedec z}%

\newcommandx\Ican[1][usedefault, addprefix=\global, 1=]{I_{#1}^{\ast}}%

\global\long\def\jsr{{\scriptstyle \mathrm{JSR}}}%

\global\long\def\cjsr{{\scriptstyle \mathrm{CJSR}}}%

\global\long\def\rsr{{\scriptstyle \mathrm{RJSR}}}%

\global\long\def\ctspc{\mathscr{M}}%

\global\long\def\b#1{\boldsymbol{#1}}%

\global\long\def\pseudy{\text{\ensuremath{\abv y}}}%

\global\long\def\regcoef{\kappa}%

\global\long\def\adj{\operatorname{adj}}%

\global\long\def\llangle{\dbllangle}%

\global\long\def\rrangle{\dblrangle}%

\global\long\def\smldblangle#1{\ensuremath{\llangle#1\rrangle}}%

\newcommand{\casecens}{{\upshape{(i)}}}

\newcommand{\caseclas}{{\upshape{(ii)}}}

\newcommand{\casestat}{{\upshape{(iii)}}}

\global\long\def\delcens{\mathrm{(i)}}%

\global\long\def\delclas{\mathrm{(ii)}}%

\global\long\def\smlf{f}%

\global\long\def\bigf{\b f}%

\global\long\def\fe{f}%

\global\long\def\ga{g}%

\global\long\def\co{\chi}%

\global\long\def\CO{\mathrm{X}}%

\global\long\def\EQ{\Theta}%

\global\long\def\eq{\theta}%

\global\long\def\ct{\psi}%

\global\long\def\cn{\mu}%

\global\long\def\cvar{{\cal M}}%

\global\long\def\codf{\chi}%

\global\long\def\eqdf{\chi_{2}}%

\global\long\def\ctdf{\chi_{1}}%

\global\long\def\set#1{\mathscr{#1}}%

\global\long\def\srp{\mathrm{H}}%

\global\long\def\srfn{h}%

\global\long\def\ch{\operatorname{co}}%

\global\long\def\im{\operatorname{im}}%

\global\long\def\lin{\mathrm{lin}}%

\global\long\def\state{\mathfrak{z}}%

\global\long\def\fespc{\mathscr{F}}%

\global\long\def\denspc{\mathscr{R}}%

\global\long\def\den{\varrho}%

\global\long\def\cden{\rho}%

\title{Inference on Common Trends \\ in a Cointegrated Nonlinear SVAR}
\author{\authaffil{James A.\ Duffy\footnotemark[1]{}}{University of Oxford}
\hspace{2cm} \authaffil{Xiyu Jiao\footnotemark[2]{}}{University
of Gothenburg}}
\date{\vspace*{0.3cm}April 2026}

\maketitle
\renewcommand*{\thefootnote}{\fnsymbol{footnote}}

\footnotetext[1]{Department\ of Economics and Corpus Christi College;
\texttt{james.duffy@economics.ox.ac.uk}}

\footnotetext[2]{Department\ of Economics; \texttt{xiyu.jiao@economics.gu.se}}

\renewcommand*{\thefootnote}{\arabic{footnote}}

\setcounter{footnote}{0}
\begin{abstract}
\noindent We consider the problem of performing inference on the number
of common stochastic trends when data is generated by a cointegrated
CKSVAR (a two-regime, piecewise affine SVAR; Mavroeidis, 2021), using
a modified version of the Breitung (2002) multivariate variance ratio
test that is robust to the presence of nonlinear cointegration (of
a known form). To derive the asymptotics of our test statistic, we
prove a fundamental LLN-type result for a class of stable but nonstationary
autoregressive processes, using a novel dual linear process approximation.
We show that our modified test yields correct inferences regarding
the number of common trends in such a system, whereas the unmodified
test tends to infer a higher number of common trends than are actually
present, when cointegrating relations are nonlinear.
\end{abstract}
\vfill

\noindent{}We thank participants at the \emph{Oxford Bulletin of
Economics and Statistics} `40 years of Unit Roots and Cointegration'
workshop, held in Oxford in April 2025, for their comments and advice.

\thispagestyle{plain}

\pagenumbering{roman}

\newpage{}

\thispagestyle{plain}

\setcounter{tocdepth}{2}

\tableofcontents{}

\newpage{}

\pagenumbering{arabic}

\newcommand{\thmcanonical}{2.1}

\global\long\def\ginv{\sim1}%

\global\long\def\dmn#1{\bar{#1}}%

\global\long\def\dtr#1{\ddot{#1}}%

\global\long\def\ctr{\tau}%

\global\long\def\breit{T}%

\global\long\def\W{\mathbb{W}}%

\global\long\def\V{\mathbb{V}}%

\global\long\def\zc{\mathfrak{z}}%

\global\long\def\trans{\top}%

\newcommand{\DMW}{\citetalias{DMW22}}

\section{Introduction}

For almost half a century, the structural vector autoregression (SVAR)
has been the workhorse model of empirical macroeconomics. In addition
to providing a tractable framework for the identification of causal
relationships in the presence of simultaneity, the model succeeds
in capturing many of the characteristic properties of macroeconomic
time series: their temporal dependence, their trending and random
wandering behaviour, and the tendency of related series to move together.
In this regard, the emergence of the theory of cointegration (\citealp{Granger86OBES};
\citealp{EG87Ecta}) was of major significance: for by formalising
that co-movement in terms of common stochastic trends, it made it
possible to identify the precise conditions under which an SVAR could
generate such common trends, as per the Granger--Johansen representation
theorem (GJRT; \citealp{Joh91Ecta,Joh95}). This result has in turn
provided the basis for a rich and fruitful theory of asymptotic inference
in cointegrated SVARs, concerning the number of common stochastic
trends in the system (or equivalently, the cointegrating rank), the
coefficients on the cointegrating relations, and the model parameters
(and implied impulse responses, etc.).

In its original conception, cointegration was inherently linear; there
have since been multifarious efforts to extend it in a nonlinear direction,
as reviewed by \citet{Tjo20EctRev}. Paralleling those efforts has
been the burgeoning of a literature on nonlinear SVARs, but which
has been confined almost entirely to the modelling of stationary time
series (see e.g.\ \citealp{Tong90}; \citealp{TTG10}; for the exceptional
case of `nonlinear VECM' models, see \citealp{KR10JoE}). This unfortunately
precludes the application of these nonlinear SVARs to settings where,
for economic reasons, the nonlinearities relate to the level of a
stochastically trending series, so that reformulating the model in
terms of the (more approximately stationary) differenced series is
not appropriate. A leading example arises in the context of the zero
lower bound (ZLB) constraint on nominal interest rates, which refers
to the level of a highly persistent -- and arguably integrated --
series, rather than to its first differences.

The development of a new class of `endogenous regime switching'
piecewise affine SVARs -- and their successful application to highly
persistent series that are subject to occasionally binding constraints
(\citealp{SM21}; \citealp{AMSV21}; \citealp{ILMZ20}) -- has recently
foregrounded the question of whether, and how, one can accommodate
stochastic trends in \emph{nonlinear} SVARs. By way of an answer,
\citet{DMW22} and \citet{DM24} provide extensions of the GJRT to
a broad class of nonlinear SVARs: in the former, to a two-regime piecewise
affine SVAR (the `CKSVAR'), and in the latter, to more general,
additively time-separable nonlinear SVARs of the form
\begin{equation}
\fe_{0}(z_{t})=c+\sum_{i=1}^{k}\fe_{i}(z_{t-i})+u_{t}\label{eq:addsep}
\end{equation}
where $z_{t}$ and $u_{t}$ are respectively the observed series and
the innovations, both of which are $\reals^{p}$-valued, and $\fe_{i}:\reals^{p}\setmap\reals^{p}$.
Their results demonstrate that, alongside linear cointegration, nonlinear
SVARs of the form \eqref{addsep} are capable of accommodating much
richer varieties of long-run behaviour than are linear SVARs, including
\emph{nonlinear} common stochastic trends and \emph{nonlinear} cointegrating
relations.

There remains the question of how to perform inference in the setting
of \eqref{addsep}, in the presence of (linear or nonlinear) cointegration.
In this paper, we consider this problem when \eqref{addsep} is specialised
to the two-regime piecewise affine model of \citet{DMW22}, as per
\begin{equation}
\phi_{0}^{+}y_{t}^{+}+\phi_{0}^{-}y_{t}^{-}+\Phi_{0}^{x}x_{t}=c+\sum_{i=1}^{k}[\phi_{i}^{+}y_{t-i}^{+}+\phi_{i}^{-}y_{t-i}^{-}+\Phi_{i}^{x}x_{t-i}]+u_{t},\label{eq:cksvar}
\end{equation}
where we have partitioned $z_{t}=(y_{t},x_{t}^{\trans})^{\trans}$
such that $y_{t}$ is $\reals$-valued and $x_{t}$ is $\reals^{p-1}$-valued,
and $y_{t}^{+}=\max\{y_{t},0\}$ and $y_{t}^{-}=\min\{y_{t},0\}$
respectively denote the positive and negative parts of $y_{t}$. We
further suppose that this model is configured such that the cointegrating
rank, $r$, is invariant to the sign of $y_{t}$, while permitting
those $r$ cointegrating relations to be nonlinear: what is termed
`case~(ii)' in the typology of \citet{DMW22}; see \secref{model}
for a discussion. Even in this case, asymptotic inference is complicated
by the fact that the processes generated by the model do not readily
fall within any class previously considered in econometrics. Although
$\{z_{t}\}$ behaves similarly, in large samples, to a (linear) integrated
process, in the sense that $n^{-1/2}z_{\smlfloor{n\lambda}}$ converges
weakly to a nondegenerate limiting process $Z(\lambda)$, neither
its first differences nor the equilibrium errors will be stationary,
but instead follow a (stable) time-varying autoregressive process,
whose coefficients depend on the sign of the integrated process $\{y_{t}\}$.
This renders any existing LLN-type results for `weakly dependent'
processes inapplicable.

In this paper we take the first steps towards the development of valid
asymptotic inference in the model \eqref{cksvar}, in the presence
of cointegration. We do so by considering the simpler problem of inference
on the cointegrating rank of \eqref{cksvar}, using a form of the
\citet{Breitung02JoE} multivariate variance ratio test statistic,
modified so as to accommodate the possibility of nonlinear cointegration.
This motivates the main technical contribution of the paper: a new
LLN-type result for the class of time-varying, stable but nonstationary
autoregressive processes that may be generated by \eqref{cksvar},
which is provided in \secref{test} along with the asymptotics of
our test statistic. This result is fundamental to the asymptotics
of estimators of the parameters of \eqref{cksvar}, the derivation
of which is the subject of the authors' ongoing research. The finite-sample
performance of our proposed test is investigated through simulation
exercises reported in \secref{finite-sample}, where it is shown that
the conventional (i.e.\ unmodified) \citet{Breitung02JoE} test tends
to incorrectly interpret the presence of \emph{nonlinear} cointegration
as evidence in favour of additional stochastic trends being present
in the data, a problem that is avoided by our proposed test. \secref{conclusion}
concludes.

\begin{notation*}
$e_{m,i}$ denotes the $i$th column of the $m\times m$ identity
matrix $I_{m}$; when $m$ is clear from the context, we write this
simply as $e_{i}$. In a statement such as $f(a^{\pm},b^{\pm})=0$,
the notation `$\pm$' signifies that both $f(a^{+},b^{+})=0$ and
$f(a^{-},b^{-})=0$ hold; similarly, `$a^{\pm}\in A$' denotes that
both $a^{+}$ and $a^{-}$ are elements of $A$. All limits are taken
as $n\goesto\infty$ unless otherwise stated. $\inprob$ and $\wkc$
respectively denote convergence in probability and in distribution
(weak convergence). We write `$X_{n}(\lambda)\wkc X(\lambda)$ on
$D_{\reals^{m}}[0,1]$' to denote that $\{X_{n}\}$ converges weakly
to $X$, where these are considered as random elements of $D_{\reals^{m}}[0,1]$,
the space of cadlag functions $[0,1]\setmap\reals^{m}$, equipped
with the uniform topology; we denote this as $D[0,1]$ whenever the
value of $m$ is clear from the context. $\smlnorm{\cdot}$ denotes
the Euclidean norm on $\reals^{m}$, and the matrix norm that it induces.
For $X$ a random vector and $p\geq1$, $\smlnorm X_{p}\defeq(\expect\smlnorm X^{p})^{1/p}$.
$C$, $C_{1}$, etc., denote generic constants that may take different
values at different places of the same proof.
\end{notation*}

\section{Model: the censored and kinked SVAR}

\label{sec:model}

\subsection{Framework}

We consider a structural VAR($k$) model in $p$ variables, in which
one series, $y_{t}$, enters with coefficients that differ according
to whether it is above or below a time-invariant threshold $b$, while
the other $p-1$ series, collected in $x_{t}$, enter linearly (\citealp{SM21};
\citealp{DMW22}). Defining
\begin{align}
y_{t}^{+} & \defeq\max\{y_{t},b\} & y_{t}^{-} & \defeq\min\{y_{t},b\},\label{eq:y-threshold}
\end{align}
we specify that $z_{t}=(y_{t},x_{t}^{\trans})^{\trans}$ follow 
\begin{equation}
\phi_{0}^{+}y_{t}^{+}+\phi_{0}^{-}y_{t}^{-}+\Phi_{0}^{x}x_{t}=c+\sum_{i=1}^{k}[\phi_{i}^{+}y_{t-i}^{+}+\phi_{i}^{-}y_{t-i}^{-}+\Phi_{i}^{x}x_{t-i}]+u_{t}\label{eq:var-two-sided}
\end{equation}
or, more compactly,
\begin{equation}
\phi^{+}(L)y_{t}^{+}+\phi^{-}(L)y_{t}^{-}+\Phi^{x}(L)x_{t}=c+u_{t},\label{eq:var-pm}
\end{equation}
where 
\begin{align*}
\phi^{\pm}(L) & \defeq\phi_{0}^{\pm}-\sum_{i=1}^{k}\phi_{i}^{\pm}L^{i} & \Phi^{x}(L) & \defeq\Phi_{0}^{x}-\sum_{i=1}^{k}\Phi_{i}^{x}L^{i},
\end{align*}
for $\phi_{i}^{\pm}\in\reals^{p\times1}$ and $\Phi_{i}^{x}\in\reals^{p\times(p-1)}$,
and $L$ denotes the lag operator. Through an appropriate redefinition
of $y_{t}$ and $c$, we may take $b$ (which we treat here as being
known) to be zero without loss of generality, and will do so throughout
the sequel. In this case, $y_{t}^{+}$ and $y_{t}^{-}$ respectively
equal the positive and negative parts of $y_{t}$, and $y_{t}=y_{t}^{+}+y_{t}^{-}$.\footnote{Throughout the following, the notation `$a^{\pm}$' connotes $a^{+}$
and $a^{-}$ as objects associated respectively with $y_{t}^{+}$
and $y_{t}^{-}$, or their lags. If we want to instead denote the
positive and negative parts of some $a\in\reals$, we shall do so
by writing $[a]_{+}\defeq\max\{a,0\}$ or $[a]_{-}\defeq\min\{a,0\}$.} Following \citet{SM21}, we term this model the `censored and kinked
SVAR' (CKSVAR), even though we here suppose that $y_{t}$ is observed
on \emph{both} sides of zero, rather than being subject to censoring.

We follow \citet{SM21} and \citet{AMSV21} in maintaining the following
conditions, which are necessary and sufficient to ensure that \eqref{var-pm}
has a unique solution for $(y_{t},x_{t})$, for all possible values
of $u_{t}$. Define 
\[
\Phi_{0}\defeq\begin{bmatrix}\phi_{0}^{+} & \phi_{0}^{-} & \Phi_{0}^{x}\end{bmatrix}=\begin{bmatrix}\phi_{0,yy}^{+} & \phi_{0,yy}^{-} & \phi_{0,yx}^{\trans}\\
\phi_{0,xy}^{+} & \phi_{0,xy}^{-} & \Phi_{0,xx}
\end{bmatrix},
\]
$\Phi_{0}^{+}\defeq[\phi_{0}^{+},\Phi_{0}^{x}]$ and $\Phi_{0}^{-}\defeq[\phi_{0}^{-},\Phi_{0}^{x}]$.

\assumpname{DGP}
\begin{assumption}
\label{ass:dgp}~
\begin{enumerate}[label=\ass{\arabic*.}, ref=\ass{.\arabic*}, itemsep=1pt,topsep=2pt]
\item \label{enu:dgp:defn} $\{(y_{t},x_{t})\}$ are generated according
to \eqref{y-threshold}--\eqref{var-pm} with $b=0$, with (possibly
random) initial values $(y_{i},x_{i})$, for $i\in\{-k+1,\ldots,0\}$;
\item \label{enu:dgp:coherence} $\sgn(\det\Phi_{0}^{+})=\sgn(\det\Phi_{0}^{-})\neq0$.
\item \label{enu:dgp:wlog} $\Phi_{0,xx}$ is invertible, and
\[
\sgn\{\phi_{0,yy}^{+}-\phi_{0,yx}^{\trans}\Phi_{0,xx}^{-1}\phi_{0,xy}^{+}\}=\sgn\{\phi_{0,yy}^{-}-\phi_{0,yx}^{\trans}\Phi_{0,xx}^{-1}\phi_{0,xy}^{-}\}>0.
\]
\item \label{enu:dgp:iid}$\{u_{t}\}_{t\in\integers}$ is an i.i.d.\ sequence
in $\reals^{p}$ with $\expect u_{t}=0$, $\expect u_{t}u_{t}^{\trans}=\Sigma_{u}$
positive definite, and $\smlnorm{u_{t}}_{2+\delta_{u}}<\infty$ for
some $\delta_{u}>0$.
\end{enumerate}
\end{assumption}

As discussed in \citet[Rem.~2.1(i)]{DMW23stat}, \ref{ass:dgp}\ref{enu:dgp:wlog}
may be maintained without loss of generality, when the invertibility
condition \ref{ass:dgp}\ref{enu:dgp:coherence} holds. Let $\{\filt_{t}\}_{t\in\integers}$
denote an underlying filtration to which the preceding processes are
all adapted. When we say that a sequence is i.i.d.,\ as per $\{u_{t}\}_{t\in\integers}$
in \assref{dgp}\ref{enu:dgp:iid}, we mean that this sequence is
$\{\filt_{t}\}_{t\in\integers}$-adapted, \emph{and additionally}
that $u_{s}$ is independent of $\filt_{t}$ for $s>t$. An immediate
implication of \assref{dgp}\ref{enu:dgp:iid} is that
\begin{equation}
U_{n}(\lambda)\defeq n^{-1/2}\sum_{t=1}^{\smlfloor{n\lambda}}u_{t}\wkc U(\lambda)\label{eq:Unwkc}
\end{equation}
on $D[0,1]$, where $U$ is a $p$-dimensional Brownian motion with
variance $\Sigma_{u}$. All the weak convergences that are stated
in this paper hold jointly with \eqref{Unwkc}.

\subsection{Canonical form}

\label{subsec:canonical}

In the terminology of \citet{DMW23stat} and \citet{DMW22}, we designate
a CKSVAR as \emph{canonical} if
\begin{equation}
\Phi_{0}=\begin{bmatrix}1 & 1 & 0\\
0 & 0 & I_{p-1}
\end{bmatrix}\eqdef\Ican[p].\label{eq:canonical}
\end{equation}
While it is not always the case that the reduced form of \eqref{var-pm}
corresponds directly to a canonical CKSVAR, by defining the canonical
variables
\begin{equation}
\begin{bmatrix}\tilde{y}_{t}^{+}\\
\tilde{y}_{t}^{-}\\
\tilde{x}_{t}
\end{bmatrix}\defeq\begin{bmatrix}\bar{\phi}_{0,yy}^{+} & 0 & 0\\
0 & \bar{\phi}_{0,yy}^{-} & 0\\
\phi_{0,xy}^{+} & \phi_{0,xy}^{-} & \Phi_{0,xx}
\end{bmatrix}\begin{bmatrix}y_{t}^{+}\\
y_{t}^{-}\\
x_{t}
\end{bmatrix}\eqdef P^{-1}\begin{bmatrix}y_{t}^{+}\\
y_{t}^{-}\\
x_{t}
\end{bmatrix},\label{eq:canon-vars}
\end{equation}
where $\bar{\phi}_{0,yy}^{\pm}\defeq\phi_{0,yy}^{\pm}-\phi_{0,yx}^{\trans}\Phi_{0,xx}^{-1}\phi_{0,xy}^{\pm}>0$
and $P^{-1}$ is invertible under \ref{ass:dgp}; and setting
\begin{equation}
\begin{bmatrix}\tilde{\phi}^{+}(\zc) & \tilde{\phi}^{-}(\zc) & \tilde{\Phi}^{x}(\zc)\end{bmatrix}\defeq Q\begin{bmatrix}\phi^{+}(\zc) & \phi^{-}(\zc) & \Phi^{x}(\zc)\end{bmatrix}P,\label{eq:canon-polys}
\end{equation}
for $\zc\in\complex$, where 
\begin{equation}
Q\defeq\begin{bmatrix}1 & -\phi_{0,yx}^{\trans}\Phi_{0,xx}^{-1}\\
0 & I_{p-1}
\end{bmatrix},\label{eq:Q-canon}
\end{equation}
we obtain a canonical CKSVAR for $\tilde{z}_{t}\defeq(\tilde{y}_{t},\tilde{x}_{t}^{\trans})^{\trans}$
(see Proposition~\thmcanonical{} in \citealp{DMW23stat}).

To distinguish between a general CKSVAR in which possibly $\Phi_{0}\neq\Ican[p]$,
and its associated canonical form, we shall refer to the former as
the `structural form' of the CKSVAR. Since the time series properties
of a general CKSVAR are largely inherited from its derived canonical
form, we shall occasionally work with this more convenient representation
of the system, and indicate this as follows.

\assumpname{DGP$^{\ast}$}
\begin{assumption}
\label{ass:dgp-canon}$\{(y_{t},x_{t})\}$ are generated by a canonical
CKSVAR, i.e.\ \assref{dgp} holds with $\Phi_{0}=[\phi_{0}^{+},\phi_{0}^{-},\Phi^{x}]=\Ican[p]$,
so that \eqref{var-two-sided} may be equivalently written as 
\begin{equation}
\begin{bmatrix}y_{t}\\
x_{t}
\end{bmatrix}=c+\sum_{i=1}^{k}\begin{bmatrix}\phi_{i}^{+} & \phi_{i}^{-} & \Phi_{i}^{x}\end{bmatrix}\begin{bmatrix}y_{t-i}^{+}\\
y_{t-i}^{-}\\
x_{t-i}
\end{bmatrix}+u_{t}.\label{eq:canon-var}
\end{equation}
\end{assumption}

\subsection{The cointegrated CKSVAR}

\citet{DMW22}, henceforth \DMW{}, develop conditions under which
the CKSVAR is capable of generating cointegrated time series. Their
work identifies three cases, which may be distinguished according
to whether stochastic trends are imparted: (i) to $y_{t}^{+}$ only
(or equivalently to $y_{t}^{-}$ only); (ii) to both $y_{t}^{+}$
and $y_{t}^{-}$; and (iii) to neither $y_{t}^{+}$ nor $y_{t}^{-}$.
Here our focus is on case~(ii), which entails that the system has
a well-defined cointegrating rank $r$, but permits the $r$ cointegrating
relationships that eliminate the ($p-r=q$) common trends to be nonlinear.
The assumptions that characterise how the model needs to be configured
for case~(ii) are given below. To state these, define the autoregressive
polynomials
\[
\Phi^{\pm}(\zc)\defeq\begin{bmatrix}\phi^{\pm}(\zc) & \Phi^{x}(\zc)\end{bmatrix},
\]
and let $\Gamma_{i}^{\pm}\defeq-\sum_{j=i+1}^{k}\Phi_{j}^{\pm}\eqdef[\gamma_{i}^{\pm},\Gamma_{i}^{x}]$
for $i\in\{1,\ldots,k-1\}$, so that $\Gamma^{\pm}(\zc)\defeq\Phi_{0}^{\pm}-\sum_{i=1}^{k-1}\Gamma_{i}^{\pm}\zc^{i}$
is such that
\[
\Phi^{\pm}(\zc)=\Phi^{\pm}(1)\zc+\Gamma^{\pm}(\zc)(1-\zc).
\]
We further define
\begin{align*}
\Pi^{\pm} & \defeq-\Phi^{\pm}(1)=-[\phi^{\pm}(1),\Phi^{x}(1)]\eqqcolon[\pi^{\pm},\Pi^{x}].
\end{align*}

\assumpname{CVAR}
\begin{assumption}
\label{ass:co}~
\begin{enumerate}[label=\ass{\arabic*.}, ref=\ass{.\arabic*}, itemsep=1pt,topsep=2pt]
\item \label{enu:co:roots}$\det\Phi^{\pm}(\zc)$ has $q^{\pm}\in\{1,\ldots,p\}$
roots at real unity, and all others outside the unit circle; and
\item \label{enu:co:rank}$\rank\Pi^{\pm}=r^{\pm}=p-q^{\pm}$.
\end{enumerate}
\end{assumption}

The preceding conditions are common to all three cases noted above.
To specialise to case~(ii), which has a constant cointegrating rank
$r=r^{+}=r^{-}$, with a stochastic trend present being in $y_{t}$,
we must additionally suppose that $\rank\Pi^{x}=r$, so that $\Pi^{\pm}$
may be written as
\[
\Pi^{\pm}=\Pi^{x}\begin{bmatrix}\theta^{\pm} & I_{p-1}\end{bmatrix}=\alpha\begin{bmatrix}\beta_{y}^{\pm} & \beta_{x}^{\trans}\end{bmatrix}\eqdef\alpha\beta^{\pm\trans},
\]
where $\alpha\in\reals^{p\times r}$, $\beta_{x}\in\reals^{(p-1)\times r}$
and $\beta^{\pm}\in\reals^{p\times r}$ have rank $r$, and $\theta^{\pm}\in\reals^{p-1}$
is such that $\Pi^{x}\theta^{\pm}=\pi^{\pm}$ (see Section~4.2 of
\DMW{}). Letting $\indic^{+}(y)\defeq\indic\{y\geq0\}$ and $\indic^{-}(y)\defeq\indic\{y<0\}$,
the (possibly nonlinear) $r$ cointegrating relationships among the
elements of $z_{t}$ are given by
\[
\beta(y)\defeq\beta^{+}\indic^{+}(y)+\beta^{-}\indic^{-}(y).
\]
Let $\alpha_{\perp}\in\reals^{p\times q}$ be such that $\alpha_{\perp}^{\trans}\alpha=0$,
and $[\alpha,\alpha_{\perp}]$ is nonsingular. The limiting form of
the stochastic trends will be a kind of (regime-dependent) projection
of the $p$-dimensional Brownian motion $U$ onto a manifold of dimension
$q=p-r$, where this projection is defined in terms of
\begin{gather}
P_{\beta_{\perp}}(y)\defeq\beta_{\perp}(y)[\alpha_{\perp}^{\trans}\Gamma(1;y)\beta_{\perp}(y)]^{-1}\alpha_{\perp}^{\trans},\label{eq:projk1}\\
\begin{aligned}\beta_{\perp}(y) & \defeq\begin{bmatrix}1 & 0\\
-\theta(y) & \beta_{x,\perp}
\end{bmatrix}, & \qquad\Gamma(1;y) & \defeq\Gamma^{+}(1)\indic^{+}(y)+\Gamma^{-}(1)\indic^{-}(y),\end{aligned}
\label{eq:betaperpfn}
\end{gather}
for $\theta(y)\defeq\indic^{+}(y)\theta^{+}+\indic^{-}(y)\theta^{-}$.
(Such objects as $P_{\beta_{\perp}}(y)$ take only two distinct values,
depending on the sign of $y$, and we routinely use the notation $P_{\beta_{\perp}}(+1)$
and $P_{\beta_{\perp}}(-1)$ to indicate these.) Define $\b{\alpha},\b{\beta}(y)\in\reals^{[k(p+1)-1]\times[r+(k-1)(p+1)]}$
as
\begin{align}
\b{\alpha}\defeq & \begin{bmatrix}\alpha & \Gamma_{1} & \Gamma_{2} & \cdots & \Gamma_{k-1}\\
 & I_{p+1}\\
 &  & I_{p+1}\\
 &  &  & \ddots\\
 &  &  &  & I_{p+1}
\end{bmatrix}, & \b{\beta}(y)^{\trans} & \defeq\begin{bmatrix}\beta(y)^{\trans}\\
S_{p}(y) & -I_{p+1}\\
 & I_{p+1} & -I_{p+1}\\
 &  & \ddots & \ddots\\
 &  &  & I_{p+1} & -I_{p+1}
\end{bmatrix},\label{eq:alpabeta}
\end{align}
where $\Gamma_{i}\defeq[\gamma_{i}^{+},\gamma_{i}^{-},\Gamma_{i}^{x}]$
for $i\in\{1,\ldots,k-1\}$, and
\begin{equation}
S_{p}(y)\defeq\begin{bmatrix}\indic^{+}(y) & \indic^{-}(y) & 0\\
0 & 0 & I_{p-1}
\end{bmatrix}^{\trans}.\label{eq:Sp}
\end{equation}
Finally, let $\rho(M)$ denote the spectral radius of $M\in\reals^{m\times m}$,
and for $\mathcal{A}\subset\reals^{m\times m}$ a bounded collection
of matrices, let
\[
\rho_{\jsr}(\mathcal{A})\defeq\limsup_{t\goesto\infty}\sup_{B\in\mathcal{A}^{t}}\rho(B)^{1/t}
\]
denote its joint spectral radius (JSR; e.g.\ \citealp{Jungers09},
Defn.\ 1.1), where $\mathcal{A}^{t}\defeq\{\prod_{s=1}^{t}M_{s}\mid M_{s}\in\mathcal{A}\}$
is the set of $t$-fold products of matrices in ${\cal A}$.

\assumpname{CO\caseclas{}}
\begin{assumption}
\label{ass:coclas}~
\begin{enumerate}[label=\ass{\arabic*.}, ref=\ass{.\arabic*},itemsep=1pt,topsep=2pt]
\item \label{enu:coclas:rank}$r^{+}=r^{-}=\rank\Pi^{x}=r$, for some $r\in\{0,1,\ldots,p-1\}$.
\item \label{enu:coclas:jsr}$\rho_{\jsr}(\{I+\tilde{\b{\beta}}(+1)^{\trans}\tilde{\b{\alpha}},I+\tilde{\b{\beta}}(-1)^{\trans}\tilde{\b{\alpha}}\})<1$.
\item \label{enu:coclas:det}$\sgn\det\alpha_{\perp}^{\trans}\Gamma(1;+1)\beta_{\perp}(+1)=\sgn\det\alpha_{\perp}^{\trans}\Gamma(1;-1)\beta_{\perp}(-1)\neq0$.
\item \label{enu:coclas:init}%
\begin{enumerate}[label=\ass{\alph*.}, ref=\ass{.\alph*}, leftmargin=0.40cm]
\item $\beta(y_{t})^{\trans}z_{t}$, and $\Delta z_{t}$ have uniformly
bounded $2+\delta_{u}$ moments, for $t\in\{-k+1,\ldots,0\}$.
\item $n^{-1/2}z_{0}\inprob{\cal Z}_{0}=[\begin{smallmatrix}{\cal Y}_{0}\\
{\cal X}_{0}
\end{smallmatrix}]$, where ${\cal Z}_{0}$ is non-random, and satisfies $\beta(\mathcal{Y}_{0})^{\trans}{\cal Z}_{0}=0$.
\end{enumerate}
\end{enumerate}
\end{assumption}

Condition \assref{coclas}\ref{enu:coclas:jsr} is stated slightly
differently from the form given in \DMW{}, so as to more directly
accommodate the case of a general (i.e.\ non-canonical) CKSVAR. In
particular, $\tilde{\b{\beta}}(y)$ and $\tilde{\b{\alpha}}$ refer
to the counterparts of \eqref{alpabeta} constructed from the parameters
of the canonical form of the CKSVAR, derived via the mapping \eqref{canon-polys}.
(So if the CKSVAR is in fact canonical, the tildes are redundant.)
See Remark~4.2(i) of \DMW{} for further details. Regarding the history
of the process prior to time $t=-k+1$, we henceforth adopt the (innocuous)
convention that 
\begin{equation}
\Delta z_{t}=0,\quad\forall t\leq-k;\label{eq:initdiff}
\end{equation}
or equivalently that $z_{t}=z_{-k}$ for all $t\leq-k$.

Finally, for the purposes of developing the asymptotics of our rank
test (\thmref{breitung} below), we shall maintain that the intercept
$c$ is such that no deterministic trends are present in any of the
model variables, as per

\assumpname{DET}
\begin{assumption}
\label{ass:det}$c\in\spn\Pi^{+}\intsect\spn\Pi^{-}$.
\end{assumption}

Under the preceding conditions (\assref{dgp}, \assref{co}, \assref{coclas}
and \assref{det}), it follows by Theorem~4.2 in \DMW{} that
\begin{equation}
n^{-1/2}\begin{bmatrix}y_{\smlfloor{n\lambda}}\\
x_{\smlfloor{n\lambda}}
\end{bmatrix}=n^{-1/2}z_{\smlfloor{n\lambda}}\wkc P_{\beta_{\perp}}[Y(\lambda)]U_{0}(\lambda)\eqdef Z(\lambda)=\begin{bmatrix}Y(\lambda)\\
X(\lambda)
\end{bmatrix},\label{eq:Zncvg}
\end{equation}
where $U_{0}(\lambda)=\Gamma(1;\mathcal{Y}_{0})\mathcal{Z}_{0}+U(\lambda)$.
(For a further heuristic discussion of the convergence in \eqref{Zncvg}
and the properties of the limiting process $Z(\lambda)$, see Section~3.3
of \DMW{}.) Since $P_{\beta_{\perp}}(\pm1)$ are rank $q$ (oblique
projection) matrices, we may regard $\{z_{t}\}$ as having $q$ common
(stochastic) trends, and $r$ cointegrating relations given by the
columns of $\beta(y)$, that eliminate those trends (since $\beta(y)^{\trans}P_{\beta_{\perp}}(y)=0$). 

On the basis of \eqref{Zncvg}, \DMW{} (see their Defn.\ 3.1) classify
$\{z_{t}\}$ as $I^{\ast}(1)$, because $n^{-1/2}z_{\smlfloor{n\lambda}}$
converges weakly to a non-degenerate process. By contrast, since the
equilibrium errors $\xi_{t}\defeq\beta(y_{t})^{\trans}z_{t}$ are
purged of the common trends in $z_{t}$, these satisfy $\max_{1\leq t\leq n}\smlnorm{\xi_{t}}=o_{p}(n^{1/2})$,
and so are of strictly smaller order than $\{z_{t}\}$; they accordingly
classify $\{\xi_{t}\}$ as $I^{\ast}(0)$. These notions of $I^{\ast}(0)$
and $I^{\ast}(1)$ processes provide a means of distinguishing between
processes whose magnitudes differ, because of the presence or absence
of stochastic trends, in a setting where the usual definitions of
$I(0)$ and $I(1)$ processes do not apply -- because in general
neither $\xi_{t}$ nor $\Delta z_{t}$ will be stationary under the
foregoing assumptions.

Although \eqref{Zncvg} implies that $Z$ is not `globally' a linear
projection of $U_{0}$ onto a $q$-dimensional linear subspace, the
following relationships hold `locally', depending on the sign of
the first component, $Y$, of $Z$:
\begin{align*}
Y(\lambda)>0 & \implies\beta^{+\trans}Z(\lambda)=0, & Y(\lambda)<0 & \implies\beta^{-\trans}Z(\lambda)=0.
\end{align*}
But in general \emph{neither} $\beta^{+\trans}Z(\lambda)$ nor $\beta^{-\trans}Z(\lambda)$
will be identically zero for all $\lambda\in[0,1]$, unless $\beta^{+}=\beta^{-}$.
The fact that there may be no rank $r=p-q$ matrix whose columns force
$Z$ to be identically zero significantly complicates the problem
of inference on the cointegrating rank, and motivates our development
of a modified form of the \citet{Breitung02JoE} test below.

\section{The modified Breitung (2002) test}

\label{sec:test}

\subsection{Fundamental ideas}

We seek to develop an (asymptotically valid) test on the cointegrating
rank $r$ -- or equivalently, the number of common trends $q$ --
that is able to accommodate the possibility of data generated by a
CKSVAR configured as per case~(ii), by adapting the approach of \citet[Sec.~5]{Breitung02JoE}.
Henceforth, as per the discussion following \eqref{y-threshold} above,
the threshold $b$ that delineates the two regimes is assumed to be
known, and normalised to zero: so that what we have denoted as $y_{t}^{+}$
and $y_{t}^{-}$ may be regarded as directly observed, rather than
depending on some prior estimator of $b$. Estimation of $b$ may
be undertaken in conjunction with the estimation of the other parameters
of the SVAR \eqref{var-two-sided}, e.g.\ by maximum likelihood,
the asymptotics of which are deferred to future work. (We anticipate
that use of a consistent estimator of $b$ would yield a test statistic
with an identical null limiting distribution to that derived below:
due to $y_{t}$ being integrated under the null, any misclassification
that results from $\hat{b}_{n}\neq b$ would affect at most $o_{p}(n^{1/2})$
observations.)

The mathematical underpinnings of \citeauthor{Breitung02JoE}'s \citeyearpar{Breitung02JoE}
test, itself a multivariate generalisation of the variance ratio test,
may be conveniently summarised as follows. (The proof of which, together
with those of all other results given in this section, appear in \appref{thmproof}.)
\begin{prop}
\label{prop:breitung}Suppose that $\{w_{n,t}\}_{t=1}^{n}$ is a triangular
array, taking values in $\reals^{d_{w}}$, such that
\begin{equation}
\frac{1}{n}\sum_{t=1}^{\smlfloor{n\lambda}}w_{n,t}\wkc\int_{0}^{\lambda}\begin{bmatrix}\W(s)\\
0_{d_{w}-\ell}
\end{bmatrix}\diff s\eqdef\begin{bmatrix}\V(\lambda)\\
0_{d_{w}-\ell}
\end{bmatrix}\label{eq:cumlcvg}
\end{equation}
on $D_{\reals^{d_{w}}}[0,1]$, where $\W$ is a random element of
$D_{\reals^{\ell}}[0,1]$ and
\begin{align}
\frac{1}{n}\sum_{t=1}^{n}w_{n,t}w_{n,t}^{\trans} & \wkc\begin{bmatrix}\int_{0}^{1}\W(s)\W(s)^{\trans}\diff s & 0\\
0 & \Omega
\end{bmatrix}\label{eq:seccvg}
\end{align}
where $\int_{0}^{1}\W(s)\W(s)^{\trans}\diff s$, $\int_{0}^{1}\V(s)\V(s)^{\trans}\diff s$
and $\Omega\in\reals^{(d_{w}-\ell)\times(d_{w}-\ell)}$ are a.s.\ positive
definite. Let $\{\lambda_{n,i}\}_{i=1}^{d_{w}}$ denote the solutions
to
\[
\det(\lambda\mathbb{B}_{n}-\mathbb{A}_{n})=0
\]
ordered as $\lambda_{n,1}\leq\lambda_{n,2}\leq\cdots\leq\lambda_{n,d_{w}}$,
for 
\begin{align*}
\mathbb{A}_{n} & \defeq\sum_{t=1}^{n}w_{n,t}w_{n,t}^{\trans}, & \mathbb{B}_{n} & \defeq\sum_{t=1}^{n}\sum_{i=1}^{t}w_{n,i}\sum_{j=1}^{t}w_{n,j}^{\trans}.
\end{align*}
Then
\begin{enumerate}
\item if $\ell_{0}=\ell$,
\[
n^{2}\sum_{i=1}^{\ell_{0}}\lambda_{n,i}\wkc\tr\left[\int_{0}^{1}\W(s)\W(s)^{\trans}\diff s\left(\int_{0}^{1}\V(s)\V(s)^{\trans}\diff s\right)^{-1}\right];
\]
\item if $\ell_{0}>\ell$, $n^{2}\sum_{i=1}^{\ell_{0}}\lambda_{n,i}\inprob\infty$.
\end{enumerate}
\end{prop}

To illustrate how \propref{breitung} provides the basis for a test
of cointegrating rank, let us suppose initially that $\{z_{t}\}$
is generated by a linear cointegrated SVAR with $q$ common trends,
or more generally by a CKSVAR satisfying the conditions above (\assref{dgp},
\assref{co}, \assref{coclas} and \assref{det}), but for which $\beta^{+}=\beta^{-}=\beta$
and $\Gamma^{+}(1)=\Gamma^{-}(1)=\Gamma(1)$. Then $P_{\beta_{\perp}}(y)$
no longer depends on (the sign of) $y$, and \eqref{Zncvg} reduces
to
\[
n^{-1/2}z_{\smlfloor{n\lambda}}\wkc\beta_{\perp}[\alpha_{\perp}^{\trans}\Gamma(1)\beta_{\perp}]^{-1}\alpha_{\perp}^{\trans}U_{0}(\lambda).
\]
It follows that by taking
\[
w_{n,t}\defeq\begin{bmatrix}n^{-1/2}\beta_{\perp}^{\trans}\\
\beta^{\trans}
\end{bmatrix}z_{t}
\]
we may linearly separate $z_{t}$ into its $q$ `integrated' (i.e.\ $I^{\ast}(1)$)
and $r=p-q$ `weakly dependent' (i.e.\ $I^{\ast}(0)$) components,
with the result that the first $q$ components of $\frac{1}{n}\sum_{t=1}^{\smlfloor{n\lambda}}w_{n,t}$
will converge weakly to a (nondegenerate) limiting process, whereas
the final $r$ components will converge to zero, exactly as in the
manner of \eqref{cumlcvg}. $\frac{1}{n}\sum_{t=1}^{n}w_{n,t}w_{n,t}^{\trans}$
then also converges to an (invertible) block diagonal matrix, as in
\eqref{seccvg}.

By \propref{breitung}, the sum of the first $q_{0}$ generalised
eigenvalues of $\mathbb{A}_{n}$ with respect to $\mathbb{B}_{n}$
will then exhibit divergent asymptotic behaviour, depending on whether
$q_{0}$ is equal to or strictly greater than $q$. This provides
the basis for the use of this quantity as a statistic for testing
hypotheses regarding the value of $q$, exactly as proposed in \citet{Breitung02JoE}.
Since these generalised eigenvalues are invariant to common linear
transformations of $\mathbb{A}_{n}$ and $\mathbb{B}_{n}$, and $w_{n,t}$
is a linear transformation of $z_{t}$, they may be computed without
knowledge of $[\beta_{\perp},\beta]$, simply by replacing each instance
of $w_{n,t}$ by $z_{t}$ in the definitions of those matrices.

\subsection{Extension to nonlinearly cointegrated series}

Suppose that we now permit $\beta^{+}\neq\beta^{-}$ and/or $\Gamma^{+}(1)\neq\Gamma^{-}(1)$.
In this case, $P_{\beta_{\perp}}(-1)$ and $P_{\beta_{\perp}}(+1)$
each have rank $q$, but may differ by a rank one matrix, and as a
result there may only be $r-1$ distinct \emph{linear} combinations
of $z_{t}$ that will be $I^{\ast}(0)$. Accordingly, applying the
usual Breitung test to $\{z_{t}\}$ directly would tend to yield the
incorrect conclusion that there are $q+1$ common trends, rather than
only $q$. (Thus for example, in a bivariate nonlinear SVAR with one
common nonlinear trend, this test may tend to conclude that there
are \emph{two} common trends and \emph{no} cointegrating relations.)

To address this problem, here we utilise the fact that the nonlinearity
in the CKSVAR is entirely a function of the sign of the first component
of $z_{t}=(y_{t},x_{t}^{\trans})^{\trans}$, such that the nonlinear
cointegrating relationships $\beta(y)$ can be rewritten as \emph{linear}
cointegrating relationships between the elements of 
\[
z_{t}^{\ast}\defeq\begin{bmatrix}y_{t}^{+}\\
y_{t}^{-}\\
x_{t}
\end{bmatrix}=\left[\begin{array}{cc}
\indic^{+}(y_{t}) & 0\\
\indic^{-}(y_{t}) & 0\\
0 & I_{p-1}
\end{array}\right]\begin{bmatrix}y_{t}\\
x_{t}
\end{bmatrix}=S_{p}(y_{t})z_{t}
\]
via
\begin{equation}
\beta(y)=\begin{bmatrix}\beta_{y}^{+\trans}\indic^{+}(y)+\beta_{y}^{-\trans}\indic^{-}(y)\\
\beta_{x}
\end{bmatrix}=\begin{bmatrix}\indic^{+}(y) & \indic^{-}(y) & 0\\
0 & 0 & I_{p-1}
\end{bmatrix}\begin{bmatrix}\beta_{y}^{+\trans}\\
\beta_{y}^{-\trans}\\
\beta_{x}
\end{bmatrix}\eqdef S_{p}(y)^{\trans}\beta^{\ast}\label{eq:beta-ast}
\end{equation}
from which it follows that
\[
\beta(y_{t})^{\trans}z_{t}=\beta^{\ast\trans}S_{p}(y_{t})z_{t}=\beta^{\ast\trans}z_{t}^{\ast}
\]
since $z_{t}^{\ast}=S_{p}(y_{t})z_{t}$; the r.h.s.\ thus gives the
$r$ linear relationships that render $\beta^{\ast\trans}z_{t}^{\ast}\distrib I^{\ast}(0)$.
As a corollary, there will be $q+1$ (linearly independent) vectors
in $\reals^{p+1}$ that extract distinct $I^{\ast}(1)$ components
from $z_{t}^{\ast}$. We obtain an additional $I^{\ast}(1)$ component,
because under case~(ii) the common trends are present in both $y_{t}^{+}$
and $y_{t}^{-}$, which appear separately as the first two components
of $z_{t}^{\ast}$.

In extracting those common trends, we are free to choose any $(q+1)$-dimensional
basis in $\reals^{p+1}$ whose span does not (non-trivially) intersect
with $\spn\beta^{\ast}$. Here we take this basis to be the columns
of the following $(p+1)\times(q+1)$ matrix
\begin{equation}
\ctr^{\ast}\defeq\begin{bmatrix}1 & 0 & \ctr_{xy}^{+\trans}\\
0 & 1 & \ctr_{xy}^{-\trans}\\
0 & 0 & \beta_{x,\perp}
\end{bmatrix},\label{eq:tau-ast}
\end{equation}
where the columns of $\beta_{x,\perp}\in\reals^{(p-1)\times(q-1)}$
span the orthogonal complement of $\spn\beta_{x}$ in $\reals^{p-1}$,
and as shown in the proof of \thmref{breitung} (see \lemref{stdised},
in particular), we are free to choose \textbf{$\ctr_{xy}^{\pm}\in\reals^{q-1}$}
so as to facilitate the convergence of our test statistic to a pivotal
limiting distribution. The matrix $\ctr^{\ast}$ plainly has rank
$q+1$; moreover the $(p+1)\times(p+1)$ matrix $[\beta^{\ast},\tau^{\ast}]$
is nonsingular, irrespective of the values of $\ctr_{xy}^{\pm}$ (see
\lemref{comp}).

Thus the linear transformation
\begin{equation}
\breit_{n}^{\trans}z_{t}^{\ast}\defeq\begin{bmatrix}n^{-1/2}\ctr^{\ast\trans}\\
\beta^{\ast\trans}
\end{bmatrix}z_{t}^{\ast}=\begin{bmatrix}n^{-1/2}\ctr^{\ast\trans}z_{t}^{\ast}\\
\beta^{\ast\trans}z_{t}^{\ast}
\end{bmatrix}=\begin{bmatrix}n^{-1/2}\ctr^{\ast\trans}z_{t}^{\ast}\\
\beta(y_{t})^{\trans}z_{t}
\end{bmatrix}\eqdef\begin{bmatrix}n^{-1/2}\varrho_{t}\\
\xi_{t}
\end{bmatrix}\eqdef\begin{bmatrix}\varrho_{n,t}\\
\xi_{t}
\end{bmatrix}\label{eq:splitting}
\end{equation}
exhaustively separates $z_{t}^{\ast}$ into its $I^{\ast}(0)$ and
(appropriately standardised) $I^{\ast}(1)$ components, and so renders
the process $\{z_{t}^{\ast}\}$ into a form conformable with \eqref{cumlcvg}
above. The decomposition \eqref{splitting} provides the basis for
applying what we term our \emph{modified Breitung} (MB) test to the
data generated by a cointegrated CKSVAR, under case~(ii), `modified'
in the sense that the test statistic will be constructed from $z_{t}^{\ast}$
rather than $z_{t}$. Indeed, if $c=0$, then it will follow from
our results below that $\frac{1}{n}\sum_{t=1}^{\smlfloor{n\lambda}}\xi_{t}\wkc0$
on $D[0,1]$, and so the test could be applied directly to $z_{t}^{\ast}$
in this case. More generally, when $c\neq0$, we need to first extract
any deterministic components whose presence would otherwise distort
the distribution of the test statistic. If we suppose that \assref{det}
holds, then no deterministic trends are present in $z_{t}$, and by
analogy with the approach taken in the linear setting, we may project
out any constant deterministic terms by applying the test not to $z_{t}^{\ast}$
but rather to 
\[
\dmn z_{t}^{\ast}\defeq z_{t}^{\ast}-\hat{\mu}_{n,z^{\ast}}
\]
where $\hat{\mu}_{n,z^{\ast}}\defeq\frac{1}{n}\sum_{t=1}^{n}z_{t}^{\ast}$,
so that now
\begin{equation}
\breit_{n}^{\trans}\dmn z_{t}^{\ast}=\breit_{n}^{\trans}(z_{t}^{\ast}-\hat{\mu}_{n,z^{\ast}})=\begin{bmatrix}\varrho_{n,t}-\hat{\mu}_{n,\varrho}\\
\xi_{t}-\hat{\mu}_{n,\xi}
\end{bmatrix}\eqdef\begin{bmatrix}\dmn{\varrho}_{n,t}\\
\dmn{\xi}_{t}
\end{bmatrix}\label{eq:demeaned}
\end{equation}
where $\hat{\mu}_{n,\xi}\defeq\frac{1}{n}\sum_{t=1}^{n}\xi_{t}$ and
$\hat{\mu}_{n,\varrho}\defeq\frac{1}{n}\sum_{t=1}^{n}\varrho_{n,t}$.

To obtain the limiting distribution of our proposed test, we shall
verify that $w_{n,t}=\breit_{n}^{\trans}\dmn z_{t}^{\ast}$ satisfies
the requirements of \propref{breitung} above. In order for \eqref{demeaned}
to conform with \eqref{cumlcvg}, we must show that
\[
\frac{1}{n}\sum_{t=1}^{\smlfloor{n\lambda}}\dmn{\xi}_{t}=\frac{1}{n}\sum_{t=1}^{\smlfloor{n\lambda}}\xi_{t}-\lambda\hat{\mu}_{n,\xi}=o_{p}(1)
\]
uniformly in $\lambda\in[0,1]$. Similarly, for the purposes of \eqref{seccvg},
require that $\frac{1}{n}\sum_{t=1}^{n}\dmn{\xi}_{t}\dmn{\xi}_{t}^{\trans}$
converges weakly to an (a.s.)\ positive definite matrix. In other
words, we require a fundamental law of large numbers (LLN) for sample
averages of the form $\frac{1}{n}\sum_{t=1}^{\smlfloor{n\lambda}}g(\xi_{t})$.
Since $\{\xi_{t}\}$ is not, in general, a stationary process, existing
results do not apply here, and this motivates the development of the
novel LLN given as \thmref{lln} below.

\subsection{LLN for regime-switching processes}

To illustrate the essential ideas, suppose for simplicity of exposition
that $k=1$, and that the CKSVAR is canonical. Then by Lemma~B.2
of \DMW{}, $\xi_{t}=\beta(y_{t})^{\trans}z_{t}$ admits the time-varying
autoregressive representation
\begin{equation}
\xi_{t}=\beta_{t}^{\trans}c+(I_{r}+\beta_{t}^{\trans}\alpha)\xi_{t-1}+\beta_{t}^{\trans}u_{t},\label{eq:xisimple}
\end{equation}
where $\{\beta_{t}\}$ is a random sequence that in general depends,
nonlinearly, on the values of $y_{t}$ and $y_{t-1}$. Under \assref{coclas}\ref{enu:coclas:jsr},
which implies that $I_{r}+\beta_{t}^{\trans}\alpha$ is drawn from
a set of matrices whose joint spectral radius is strictly bounded
by unity, $\{\xi_{t}\}$ will be a `stable' process in the sense
that it is stochastically bounded; but the dependence of $\beta_{t}$
on $y_{t}$ prevents $\{\xi_{t}\}$ from being stationary.

Since $\beta_{t}=\beta^{+}$ whenever $y_{t-1}>0$ and $y_{t}>0$,
it follows that if $y_{s}>0$ for all $s\in\{t-m,\ldots,t\}$, then
\[
\xi_{t}=(I_{r}+\beta^{+\trans}\alpha)^{m}\xi_{t-m}+\sum_{\ell=0}^{m-1}(I_{r}+\beta^{+\trans}\alpha)^{\ell}\beta^{+\trans}(c+u_{t-\ell}).
\]
Since $\{y_{t}\}$ has a stochastic trend, it will tend to make lengthy
sojourns above the origin, during which periods $\xi_{t}$ will be
well approximated by the stationary linear process,
\[
\xi_{t}^{+}\defeq-(\beta^{+\trans}\alpha)^{-1}\beta^{+\trans}c+\sum_{\ell=0}^{\infty}(I_{r}+\beta^{+\trans}\alpha)^{\ell}\beta^{+\trans}u_{t-\ell}\eqdef\mu_{\xi}^{+}+w_{t}^{+}
\]
On the other hand, $\{y_{t}\}$ will also tend to spend lengthy epochs
below the origin, permitting $\xi_{t}$ to then be approximated by
\[
\xi_{t}^{-}\defeq-(\beta^{-\trans}\alpha)^{-1}\beta^{-\trans}c+\sum_{\ell=0}^{\infty}(I_{r}+\beta^{-\trans}\alpha)^{\ell}\beta^{-\trans}u_{t-\ell}\eqdef\mu_{\xi}^{-}+w_{t}^{-}.
\]

This reasoning suggests a kind of `dual linear process' approximation
to $\xi_{t}$, leading to an argument along the lines of
\begin{align*}
\frac{1}{n}\sum_{t=1}^{\smlfloor{n\lambda}}g(\xi_{t})\indic^{+}(y_{t}) & =\frac{1}{n}\sum_{t=1}^{\smlfloor{n\lambda}}g(\xi_{t}^{+})\indic^{+}(y_{t})+o_{p}(1)\\
 & =[\expect g(\xi_{0}^{+})]\frac{1}{n}\sum_{t=1}^{\smlfloor{n\lambda}}\indic^{+}(y_{t})+o_{p}(1)\\
 & \wkc[\expect g(\xi_{0}^{+})]\int_{0}^{\lambda}\indic^{+}[Y(s)]\diff s\eqqcolon[\expect g(\xi_{0}^{+})]m_{Y}^{+}(\lambda)
\end{align*}
where $m_{Y}^{+}(\lambda)$ measures the fraction of the interval
$[0,\lambda]$ for which $Y(s)\geq0$. We thus arrive at
\begin{align*}
\frac{1}{n}\sum_{t=1}^{\smlfloor{n\lambda}}g(\xi_{t}) & =\frac{1}{n}\sum_{t=1}^{\smlfloor{n\lambda}}g(\xi_{t})[\indic^{+}(y_{t})+\indic^{-}(y_{t})]\wkc[\expect g(\xi_{0}^{+})]m_{Y}^{+}(\lambda)+[\expect g(\xi_{0}^{-})]m_{Y}^{-}(\lambda),
\end{align*}
which will in general be random (so that the convergence is merely
in distribution), except in the special case where $\expect g(\xi_{0}^{+})=\expect g(\xi_{0}^{-})=\mu_{g}$
-- whereupon the r.h.s.\ collapses to $\lambda\mu_{g}$, since $m_{Y}^{+}(\lambda)+m_{Y}^{-}(\lambda)=\lambda$.
(Importantly for the purposes of our test, such a case systematically
arises under our assumptions, when $g(\xi)=\xi$.) The randomness
of the limit provides another manifestation of the non-ergodicity
of $\{\xi_{t}\}$, induced as by the dependence of its law of motion
on the level of $y_{t}$.

Such arguments, in the more general setting of a (not necessarily
canonical) CKSVAR($k$), lead to the main technical contribution of
this paper, a LLN-type result for additive functionals of a class
of time-varying autoregressive processes, of which \eqref{xisimple}
is a special case. To facilitate its use in other contexts, we prove
this result supposing that the following weaker condition holds in
place of \assref{det}.

\assumpname{DET$^{\prime}$}
\begin{assumption}
\label{ass:detprime}$e_{1}^{\trans}P_{\beta_{\perp}}(+1)c=0$.
\end{assumption}

The preceding permits the model to impart deterministic trends to
$x_{t}$ (but not to $y_{t}$), and leads us to consider the linearly
detrended process
\[
\begin{bmatrix}y_{t}^{d}\\
x_{t}^{d}
\end{bmatrix}=z_{t}^{d}\defeq z_{t}-[P_{\beta_{\perp}}(+1)c]t,\quad t\geq1
\]
in place of $z_{t}$, with the convention that $z_{t}^{d}\defeq z_{t}$
for $t\leq0$; note that $y_{t}^{d}=y_{t}$ (see Section~4.4 in \DMW{}).
Recall that, as per the remarks following the statement of \assref{dgp}
above, there is an underlying filtration $\{\filt_{t}\}_{t\in\integers}$
to which $\{u_{t}\}$ and $\{z_{t}\}$ are adapted, and that an i.i.d.\ process
$\{v_{t}\}$ is one that is both $\filt_{t}$-adapted, and such that
$v_{s}$ is independent of $\filt_{t}$ for $s>t$.
\begin{thm}
\label{thm:lln}Suppose \assref{dgp}, \assref{co}, \assref{coclas}
and \assref{detprime} hold. Let $\{A_{t}\}$, $\{B_{t}\}$ and $\{c_{t}\}$
be random sequences adapted to $\{\filt_{t}\}$, respectively taking
values in $\reals^{d_{w}\times d_{w}}$, $\reals^{d_{w}\times d_{v}}$
and $\reals^{d_{w}}$, where $t\in\integers$. Suppose $\{v_{t}\}$
is i.i.d\ with $\expect v_{t}=0$, and that $\{w_{t}\}$ satisfies
\begin{equation}
w_{t}=c_{t}+A_{t}w_{t-1}+B_{t}v_{t}\label{eq:wproc}
\end{equation}
for $t\geq-k$ and some given (random) $w_{-k}$ (with $w_{t}\defeq0$
for all $t\leq-k-1$); and:
\begin{enumerate}[itemsep=2pt,topsep=3pt]
\item $A_{t}\in\mset A$, $B_{t}\in\mset B$ and $c_{t}\in{\cal C}$ for
all $t\in\naturals$, where $\mset A$, $\mset B$ and ${\cal C}$
are bounded subsets of $\reals^{d_{w}\times d_{w}}$, $\reals^{d_{w}\times d_{v}}$
and $\reals^{d_{w}}$ respectively, and $\rho_{\jsr}(\mset A)<1$;
\item there exist $A^{\pm}\in\mset A$, $B^{\pm}\in\mset B$ and $c^{\pm}\in\mset C$
such that
\begin{align*}
y_{t-1}>0\text{ and }y_{t}>0 & \implies A_{t}=A^{+}\sep B_{t}=B^{+}\sep c_{t}=c^{+},\\
y_{t-1}<0\text{ and }y_{t}<0 & \implies A_{t}=A^{-}\sep B_{t}=B^{-}\sep c_{t}=c^{-};
\end{align*}
\item $m_{0}\geq1$ is such that $\smlnorm{w_{0}}_{m_{0}}+\smlnorm{v_{0}}_{m_{0}}<\infty$.
\item \label{enu:pmavg:jac}$g:\reals^{d_{w}}\setmap\reals^{d_{g}}$ is
a continuous function satisfying
\begin{equation}
\smlnorm{g(w)-g(w^{\prime})}\leq C(1+\smlnorm w^{\ell_{0}}+\smlnorm{w^{\prime}}^{\ell_{0}})\smlnorm{w-w^{\prime}}\label{eq:lip}
\end{equation}
for all $w,w^{\prime}\in\reals^{d_{w}}$, for some $0\leq\ell_{0}<m_{0}-1$.
\end{enumerate}
Then $\expect\smlnorm{g(w_{0}^{\pm})}<\infty$, and on $D[0,1]$,
\begin{equation}
\frac{1}{n}\sum_{t=1}^{\smlfloor{n\lambda}}g(w_{t})\indic^{\pm}(y_{t})\wkc[\expect g(w_{0}^{\pm})]\int_{0}^{\lambda}\indic^{\pm}[Y(\mu)]\diff\mu,\label{eq:gavg}
\end{equation}
where
\begin{equation}
w_{0}^{\pm}=(I_{d_{w}}-A^{\pm})^{-1}c^{\pm}+\sum_{\ell=0}^{\infty}(A^{\pm})^{\ell}B^{\pm}v_{-\ell}.\label{eq:w0}
\end{equation}
Moreover, 
\begin{equation}
\frac{1}{n^{3/2}}\sum_{t=1}^{\smlfloor{n\lambda}}[g(w_{t})\otimes z_{t}^{d}]\indic^{\pm}(y_{t})\wkc[\expect g(w_{0}^{\pm})]\otimes\int_{0}^{\lambda}Z(\mu)\indic^{\pm}[Y(\mu)]\diff\mu,\label{eq:gzavg}
\end{equation}
jointly with $U_{n}\wkc U$.
\end{thm}

\subsection{Limiting distribution and consistency}

Using \thmref{lln} and the representation theory of \DMW{}, we are
able to derive the limiting distribution of our modified Breitung
(MB) statistic for testing the null of $q_{0}$ common trends (and
$r_{0}=p-q_{0}$ cointegrating relations), which is defined as
\begin{equation}
\Lambda_{n,q_{0}}\defeq n^{2}\sum_{i=1}^{q_{0}+1}\lambda_{n,i}\label{eq:Lambdanq}
\end{equation}
where $\{\lambda_{n,i}\}_{i=1}^{p+1}$ are the solutions to
\begin{equation}
\det(\lambda\mathbf{B}_{n}-\mathbf{A}_{n})=0\label{eq:det-actual}
\end{equation}
ordered as $\lambda_{n,1}\leq\lambda_{n,2}\leq\cdots\leq\lambda_{n,p+1}$,
for
\begin{align}
\mathbf{A}_{n} & \defeq\sum_{t=1}^{n}\dmn z_{t}^{\ast}\dmn z_{t}^{\ast\trans}, & \mathbf{B}_{n} & \defeq\sum_{t=1}^{n}\sum_{i=1}^{t}\dmn z_{i}^{\ast}\sum_{j=1}^{t}\dmn z_{j}^{\ast\trans}.\label{eq:AnBn-actual}
\end{align}
This statistic has the same form as that considered in \propref{breitung},
though note that for testing the null of $q_{0}$ common trends we
sum over the first $q_{0}+1$ generalised eigenvalues $\{\lambda_{n,i}\}_{i=1}^{q_{0}+1}$,
reflecting the fact that $y_{t}^{+}$ and $y_{t}^{-}$ separately
enter $z_{t}^{\ast}$.

To state the limiting distribution of the test statistic, define 
\begin{equation}
W_{0}(\lambda)\defeq{\cal W}_{0}e_{q,1}+W(\lambda),\label{eq:stdbm}
\end{equation}
where ${\cal W}_{0}\in\reals$ is nonrandom, and $W$ is a $q$-dimensional
standard Brownian motion. Define the $(q+1)$-dimensional process
\begin{equation}
W_{0}^{\ast}(\lambda)\defeq S_{q}[e_{1}^{\trans}W_{0}(\lambda)]W_{0}(\lambda)\eqqcolon\begin{bmatrix}[W_{0,1}(\lambda)]_{+}\\{}
[W_{0,1}(\lambda)]_{-}\\
W_{0,-1}(\lambda)
\end{bmatrix}\label{eq:W0ast}
\end{equation}
and define $\dmn W_{0}^{\ast}(\lambda)$ to be the residual from the
pathwise $L^{2}[0,1]$ projection of each element of $W_{0}^{\ast}$
onto a constant. Let $\dmn V_{0}^{\ast}(\lambda)\defeq\int_{0}^{\lambda}\dmn W_{0}^{\ast}(\mu)\diff\mu$
denote the cumulation of $\dmn W_{0}^{\ast}$.

We only provide limit theory here for the case where $y_{0}=o_{p}(n^{1/2})$.
This simplifies the asymptotics of the testing problems in two respects:
(i) it ensures that the limiting process visits both regimes (positive
and negative) with probability one, so that the relevant matrices
are positive definite a.s.; (ii) it yields a distribution for the
test statistic that (upon demeaning) is nuisance parameter free, being
invariant to $X(0)=\mathcal{X}_{0}$. (Possible extensions to handle
the case where $n^{-1/2}y_{0}\inprob\mathcal{Y}_{0}\neq0$ are discussed
below.) In the following statement, $q$ denotes the actual (i.e.\ the
true) number of common trends in the system, whereas $q_{0}$ denotes
the null hypothesised value, i.e.\ the number used to compute the
test statistic.
\begin{thm}
\label{thm:breitung}Suppose \assref{dgp}, \assref{co}, \assref{coclas}
and \assref{det} hold, with $y_{0}=o_{p}(n^{1/2})$. Then for $W_{0}$
as defined in \eqref{stdbm}, with ${\cal W}_{0}=0$:
\begin{enumerate}
\item \label{enu:qeqq0}if $q_{0}=q$,
\begin{equation}
\Lambda_{n,q_{0}}=\Lambda_{n,q}\wkc\tr\left[\int_{0}^{1}\dmn W_{0}^{\ast}(s)\dmn W_{0}^{\ast}(s)^{\trans}\diff s\left(\int_{0}^{1}\dmn V_{0}^{\ast}(s)\dmn V_{0}^{\ast}(s)^{\trans}\diff s\right)^{-1}\right]\eqdef\Lambda_{q}\label{eq:Lambdalim}
\end{equation}
\item \label{enu:q0ltq}if $q_{0}<q$, the weak limit of $\Lambda_{n,q_{0}}$
is stochastically dominated by $\Lambda_{q}$; and
\item \label{enu:q0gtq}if $q_{0}>q$, $\Lambda_{n,q_{0}}\inprob\infty$.
\end{enumerate}
Moreover, the convergence in \eqref{Lambdalim} holds jointly with
$U_{n}\wkc U$, and with
\begin{equation}
n^{-1/2}y_{\smlfloor{n\lambda}}\wkc Y(\lambda)=\omega^{+}[e_{1}^{\trans}W_{0}(\lambda)]_{+}+\omega^{-}[e_{1}^{\trans}W_{0}(\lambda)]_{-},\label{eq:YW0}
\end{equation}
where the latter convergence also holds if $n^{-1/2}y_{0}\inprob\mathcal{Y}_{0}$
with $\mathcal{Y}_{0}$ possibly nonzero.
\end{thm}

Part~(i) of the preceding implies that valid asymptotic critical
values for $H_{0}:q=q_{0}$ can be drawn from the distribution of
$\Lambda_{q_{0}}$ (which equals $\Lambda_{q}$ under $H_{0}$); these
may be computed by simulation. Part~(ii) implies that $\Lambda_{n,q_{0}}$
is stochastically bounded when the true number of common trends ($q$)
is greater than the hypothesised number ($q_{0}$), such that a test
of $H_{0}:q=q_{0}$ will not be consistent against the alternative
$H_{1}:q>q_{0}$. On the other hand, by part~(iii), it will be consistent
against $H_{1}:q<q_{0}$. This suggests that the estimation of $q$
may be effected via a stepwise testing procedure, starting with the
null $H_{0}:q=p$ of no cointegration, and progressing downwards (i.e.\ testing
$H_{0}:q=p-1$ if the preceding null is rejected, etc., and stopping
at the first $q_{0}$ for which $H_{0}:q=q_{0}$ is not rejected).

\subsection{Extensions}

Once we allow that $n^{-1/2}y_{0}\inprob\mathcal{Y}_{0}$, with $\mathcal{Y}_{0}$
possibly nonzero, the preceding runs into certain difficulties. If
$\mathcal{Y}_{0}=0$, then ${\cal W}_{0}=0$ also, and so $W_{0,1}$
visits both sides of the origin at some point during $[0,1]$ (indeed,
during any subinterval $[0,\lambda]$) with probability one. But if
$\mathcal{Y}_{0}\neq0$ then ${\cal W}_{0}\neq0$, and this event
is no longer guaranteed to occur, with the consequence that $\int\dmn W_{0}^{\ast}\dmn W_{0}^{\ast\trans}$
and $\int\dmn V_{0}^{\ast}\dmn V_{0}^{\ast\trans}$ are no longer
positive definite with probability one. In a sense, this is merely
a technical rather than a practical problem, because the failure of
$W_{0,1}$ to visit both sides of the origin is the large-sample counterpart
of the possibility that $\{y_{t}\}$ itself may not visit both sides
of the origin either; and were it to fail to do so, the observed data
would be well (indeed, perfectly) approximated by a linearly cointegrated
system, with cointegrating relations given by either $\beta^{+}$
or $\beta^{-}$ (depending on whether $\{y_{t}\}_{t=1}^{n}$ was always
positive or negative, respectively).

The fact that we would only contemplate conducting (the modified version
of) the test in cases where $\{y_{t}\}$ spends an appreciable amount
of time in both regimes also suggests a remedy for this problem. Namely,
that we should refer the test statistic $\Lambda_{n,q_{0}}$ not to
the quantiles of its unconditional limiting distribution, but to those
of its distribution conditional on $\{y_{t}\}$ (and therefore $W_{0,1}$)
spending more than a certain fraction of the sample in each regime;
this thereby avoids the rank deficiency problem. That is, letting
${\cal M}\defeq\min\{m_{W_{0,1}}^{+}(1),m_{W_{0,1}}^{-}(1)\}$, we
propose to compare $\Lambda_{n,q_{0}}$ with the $1-\alpha$ quantile
of the distribution of $\Lambda_{q_{0}}$ conditional on ${\cal M}\geq\tau$,
i.e.\ choosing a critical value $c_{\alpha,1}(\tau)$ such that 
\begin{equation}
\Prob\{\Lambda_{q_{0}}\geq c_{\alpha,1}(\tau)\mid{\cal M}\geq\tau\}=\alpha\label{eq:condidist}
\end{equation}
where $\tau\in(0,0.5)$ is some user-specified value (say, ten or
fifteen percent).

The preceding remains well defined when $\mathcal{Y}_{0}\neq0$, but
in that case the (conditional) distribution of $\Lambda_{q_{0}}$
will depend on the unknown nuisance parameter $\mathcal{W}_{0}$.
Since the sign of $y_{0}$ and therefore $Y(0)=\mathcal{Y}_{0}$ is
known, $\mathcal{W}_{0}$ may be estimated when (say) $y_{0}>0$ on
the basis of the representation \eqref{YW0} as $(\hat{\omega}_{n}^{+})^{-1}(n^{-1/2}y_{0})$,
where
\begin{align*}
\hat{\omega}_{n}^{+} & \defeq\left(\sum_{\ell=-L_{n}}^{L_{n}}K(\ell/L_{n})\hat{\gamma}_{\ell}^{+}\right)^{1/2} & \hat{\gamma}_{\ell}^{+} & \defeq\frac{1}{\sum_{t=1}^{n}\indic^{+}(y_{t})}\sum_{t=\ell+1}^{n}\Delta y_{t}\Delta y_{t-\ell}\indic^{+}(y_{t})
\end{align*}
denotes a long-run variance estimator, with kernel $K$ and lag truncation
sequence $L_{n}\goesto\infty$. (If on the other hand $y_{0}<0$,
then an estimator $\hat{\omega}_{n}^{-}$ of $\omega^{-}$ would be
constructed analogously.)

\section{Finite-sample performance}

\label{sec:finite-sample}

Here we report the results of Monte Carlo simulations conducted to
evaluate the performance of the proposed test. We generate data from
a bivariate (i.e.\ $p=2$) cointegrated CKSVAR with $q=1$ common
trends (and so $r=1$ cointegration relations),
\[
\begin{bmatrix}\Delta y_{t}\\
\Delta x_{t}
\end{bmatrix}=c+\alpha\beta^{\ast\trans}\begin{bmatrix}y_{t-1}^{+}\\
y_{t-1}^{-}\\
x_{t-1}
\end{bmatrix}+u_{t},
\]
where $\alpha=(0.5,0.1)^{\trans}$, $\beta^{\ast}=(\beta_{y}^{+},\beta_{y}^{-},1)^{\trans}$,
$c=2\alpha$, $z_{0}=(y_{0},x_{0})^{\trans}=0$ and $u_{t}\distiid N[0,I_{2}]$.
We set $\beta_{y}^{+}=-1$, and consider a \emph{linear} design in
which $\beta_{y}^{-}=-1$, and a \emph{nonlinear} design in which
$\beta_{y}^{-}=-0.5$. The implied cointegrating vectors are $\beta^{+}=\beta^{-}=(-1,1)^{\trans}$
in the former, and $\beta^{+}=(-1,1)^{\trans}$ and $\beta^{-}=(-0.5,1)^{\trans}$
in the latter. In both cases, the assumptions of \thmref{breitung}
are satisfied; for example it may be verified that $\smlabs{1+\beta^{\pm\trans}\alpha}<1$,
so that the stability condition \assref{coclas}\ref{enu:coclas:jsr}
holds. The sample size ranges over $n\in\{200,500,1000,1500\}$. We
only retain samples in which $\{y_{t}\}$ spends at least $0.15n$
observations both above and below zero.

For each dataset thus generated, we test the null that $H_{0}:q=q_{0}$
using the following test statistics:
\begin{enumerate}
\item The \emph{standard} \emph{Breitung} (SB) test is that given in \citet[Sec.~5]{Breitung02JoE}.
In this case, $\mathbf{A}_{n}$ and $\mathbf{B}_{n}$ in \eqref{AnBn-actual}
are computed on the basis of $\dmn z_{t}$, rather than $\dmn z_{t}^{\ast}$;
\item The \emph{modified} Breitung (MB) test is our proposed test statistic,
based on $\dmn z_{t}^{\ast}$, and using a `partially conditional'
critical value $c_{\alpha,1}(\tau)$ as in \eqref{condidist} with
$\tau=0.15$.
\end{enumerate}
(Note that to test the null that $H_{0}:q=q_{0}$, SB sums over the
first $q_{0}$ generalised eigenvalues of a $p$-dimensional system,
whereas MB sums over the first $q_{0}+1$ generalised eigenvalues
of a $(p+1)$-dimensional system.) Let $q$ denote the true number
of common trends. Since the true number of common trends $q=1$ in
the foregoing designs, we test $H_{0}:q=1$ to evaluate size and $H_{0}:q=2$
to evaluate power, with a nominal significance level of $10$~per
cent. (We run 10000 Monte Carlo replications for every design.)

\begin{table}
\begin{centering}
\begin{tabular}{lcccccccccccc}
\toprule 
\addlinespace
Design & \multicolumn{4}{c}{Linear } &  & \multicolumn{4}{c}{Nonlinear } &  & \multicolumn{2}{c}{Stationary}\tabularnewline
 & \multicolumn{4}{c}{$(\beta^{+}=\beta^{-}$, $q=1$)} &  & \multicolumn{4}{c}{$(\beta^{+}\neq\beta^{-}$, $q=1$)} &  & \multicolumn{2}{c}{($q=0$)}\tabularnewline\addlinespace
\midrule
\addlinespace
$H_{0}:$ & \multicolumn{2}{c}{$q=1$} & \multicolumn{2}{c}{$q=2$} &  & \multicolumn{2}{c}{$q=1$} & \multicolumn{2}{c}{$q=2$} &  & \multicolumn{2}{c}{$q=1$}\tabularnewline\addlinespace
$n$ & \multicolumn{1}{c}{SB} & \multicolumn{1}{c}{MB} & \multicolumn{1}{c}{SB} & \multicolumn{1}{c}{MB} &  & \multicolumn{1}{c}{SB} & \multicolumn{1}{c}{MB} & \multicolumn{1}{c}{SB} & MB &  & SB & MB\tabularnewline\addlinespace
\midrule
\addlinespace
200 & 0.09 & 0.06 & 0.94 & 0.68 &  & 0.06 & 0.02 & 0.57 & 0.36 &  & 0.40 & 0.38\tabularnewline\addlinespace
500 & 0.09 & 0.09 & 1.00 & 0.95 &  & 0.08 & 0.05 & 0.64 & 0.75 &  & 0.71 & 0.81\tabularnewline\addlinespace
1000 & 0.10 & 0.10 & 1.00 & 1.00 &  & 0.08 & 0.08 & 0.61 & 0.94 &  & 0.91 & 0.98\tabularnewline\addlinespace
1500 & 0.10 & 0.10 & 1.00 & 1.00 &  & 0.08 & 0.08 & 0.58 & 0.98 &  & 0.97 & 1.00\tabularnewline\addlinespace
\bottomrule
\end{tabular}
\par\end{centering}
\caption{Rejection rates; nominal level 10 per cent}

\label{tbl:simulation}
\end{table}

The results are displayed in the first eight columns of \tblref{simulation}.
In line with our expectations, the standard Breitung test performs
poorly in the nonlinear design, having a noticeable tendency to incorrectly
find that $q=2$. This problem is remedied by the modified Breitung
test, at least for sufficiently large sample sizes, at the cost of
the test being somewhat conservative in small samples. Both tests
appear to be approximately correctly sized for testing $H_{0}:q=1$,
and both (as expected) perform well in the linear design.

As an additional check on the performance of these tests, per a request
from a referee, we also evaluated their power to reject $H_{0}:q=1$
using data generated under the following \emph{stationary} ($q=0$)
nonlinear design,
\[
\begin{bmatrix}\Delta y_{t}\\
\Delta x_{t}
\end{bmatrix}=c+\begin{bmatrix}\pi^{+} & \pi^{-} & \Pi^{x}\end{bmatrix}\begin{bmatrix}y_{t-1}^{+}\\
y_{t-1}^{-}\\
x_{t-1}
\end{bmatrix}+u_{t}=\begin{bmatrix}-0.2 & -0.5 & \phantom{-}0.3\\
\phantom{-}0.1 & \phantom{-}0.1 & -0.2
\end{bmatrix}\begin{bmatrix}y_{t-1}^{+}\\
y_{t-1}^{-}\\
x_{t-1}
\end{bmatrix}+u_{t}.
\]
The results are reported in the final two columns of \tblref{simulation},
and show that both the SB and MB tests have substantial power in this
direction. Since the `cointegrating space' is $\{0\}$ in this case,
and so trivially linear, the similar performance of the two tests
is not surprising.

\section{Conclusion}

\label{sec:conclusion}

This paper has considered the problem of testing the cointegrating
rank in a CKSVAR, proposing a modified version of the \citet{Breitung02JoE}
test that is robust to the forms of nonlinear cointegration that may
be generated by that model. En route to deriving the asymptotics of
this test, we have proved a novel LLN-type result for a class of stable
but nonstationary autoregressive processes. This result underpins
the development of the asymptotics of likelihood-based estimators
of the cointegrated CKSVAR, our results on which will be reported
elsewhere.

{\singlespacing

\bibliographystyle{ecta}
\bibliography{cksvar}

}

\appendix

\section{Auxiliary lemmas}

We here collect the fundamental technical results that are needed
for the proof of Theorems~\ref{thm:lln} and \ref{thm:breitung}.
These are all stated for a CKSVAR in canonical form, i.e.\ supposing
that \assref{dgp-canon} holds. For a general CKSVAR, i.e.\ one satisfying
\assref{dgp} rather than \assref{dgp-canon}, Proposition~2.1 in
\DMW{} establishes that there is a linear mapping between $z_{t}^{\ast}$
and a derived canonical process $\tilde{z}_{t}^{\ast}$ satisfying
\assref{dgp-canon}. Because $\Lambda_{n,q}$ is invariant to (common)
linear transformations of $\mathbf{A}_{n}$ and $\mathbf{B}_{n}$,
as defined in \eqref{AnBn-actual}, the asymptotics of the canonical
process accordingly govern the large-sample behaviour of our test
statistic.

We first recall that under \assref{dgp-canon}, \assref{co}, \assref{coclas}
and \assref{detprime}, it follows by Theorems~4.2 and 4.4 of \DMW{}
that
\begin{equation}
n^{-1/2}\begin{bmatrix}y_{\smlfloor{n\lambda}}\\
x_{\smlfloor{n\lambda}}^{d}
\end{bmatrix}=n^{-1/2}z_{\smlfloor{n\lambda}}^{d}\wkc P_{\beta_{\perp}}[Y(\lambda)]U_{0}(\lambda)=\begin{bmatrix}Y(\lambda)\\
X(\lambda)
\end{bmatrix}=Z(\lambda)\label{eq:zd-cvg}
\end{equation}
on $D[0,1]$ with the further implication (via Lemma~B.3 of \DMW{})
that 
\begin{equation}
Y(\lambda)=h[\vartheta^{\trans}U_{0}(\lambda)]\vartheta^{\trans}U_{0}(\lambda)\label{eq:Ylim}
\end{equation}
where $h(u)=\indic^{+}(u)h^{+}+\indic^{-}(u)h^{-}$ for $h^{+}=1$
and $h^{-}>0$, and $\vartheta^{\trans}\defeq e^{\trans}P_{\beta_{\perp}}(+1)\neq0$.
We note that as a consequence of \eqref{zd-cvg}, \assref{coclas}\ref{enu:coclas:init}
and our (innocuous) convention that $\Delta z_{i}=0$ for $i\leq-k$
(as per \eqref{initdiff} above) that
\begin{align}
n^{-1/2}\sup_{s\leq n}\smlnorm{z_{s}^{d}} & =O_{p}(1), & n^{-1/2}\sup_{s\leq n}\smlnorm{\Delta z_{s}^{d}} & =o_{p}(1).\label{eq:supbnd}
\end{align}
Indeed, it follows by Lemmas~A.1 and B.2 of \DMW{} that
\begin{equation}
\sup_{s\in\integers}\smlnorm{\Delta z_{s}^{d}}_{2+\delta_{u}}<\infty.\label{eq:dzdmom}
\end{equation}
(Recall that for $X$ a random vector, and $p\geq1$, $\smlnorm X_{p}\defeq(\expect\smlnorm X^{p})^{1/p}$.)

\begin{lem}
\label{lem:YZwkc}Suppose \assref{dgp-canon}, \assref{co}, \assref{coclas}
and \assref{detprime} hold. Then:
\begin{enumerate}
\item \label{enu:YZwkc:ydelta}as $n\goesto\infty$ and then $\delta\goesto0$
\[
\frac{1}{n}\sum_{t=1}^{n}\indic\{n^{-1/2}\smlabs{y_{t}}\leq\delta\}\inprob0;
\]
\item \label{enu:YZwkc:YZ}on $D[0,1]$ jointly with $U_{n}\wkc U$, 
\[
\frac{1}{n}\sum_{t=1}^{\smlfloor{n\lambda}}\indic^{\pm}(y_{t})\begin{bmatrix}1\\
n^{-1/2}z_{t}^{d}
\end{bmatrix}\wkc\int_{0}^{\lambda}\indic^{\pm}[Y(\mu)]\begin{bmatrix}1\\
Z(\mu)
\end{bmatrix}\diff\mu.
\]
\end{enumerate}
\end{lem}

The following is a slightly restricted counterpart of \thmref{lln},
which holds under \assref{dgp-canon} rather than \assref{dgp}. It
will in turn be used to prove \thmref{lln} in \appref{thmproof}.
\begin{lem}
\label{lem:pmavg}Suppose \assref{dgp-canon}, \assref{co}, \assref{coclas}
and \ref{ass:detprime} hold. Then the conclusions of \thmref{lln}
hold.
\end{lem}

For the next two results, we specialise from \assref{detprime} to
\assref{det}, so that no deterministic trends are present in any
components of $z_{t}$, which is identically equal to $z_{t}^{d}$.
Recall the definitions of $\dmn{\varrho}_{n,t}$ and $\dmn{\xi}_{t}$
given in \eqref{demeaned}. We note also that as an immediate consequence
of \eqref{zd-cvg} and the continuous mapping theorem, on $D[0,1]$,
\begin{equation}
n^{-1/2}z_{\smlfloor{n\lambda}}^{\ast}=S_{p}(n^{-1/2}y_{\smlfloor{n\lambda}})n^{-1/2}z_{\smlfloor{n\lambda}}\wkc S_{p}[Y(\lambda)]Z(\lambda)\eqdef Z^{\ast}(\lambda)\label{eq:zastcvg}
\end{equation}
for $S_{p}(y)$ as in \eqref{Sp}, and hence
\begin{equation}
\varrho_{n,\smlfloor{n\lambda}}=\ctr^{\ast\trans}n^{-1/2}z_{\smlfloor{n\lambda}}^{\ast}\wkc\ctr^{\ast\trans}Z^{\ast}(\lambda)\eqdef R(\lambda),\label{eq:rhocvg}
\end{equation}
for $\tau^{\ast}$ as in \eqref{tau-ast}. Since $z_{t}^{\ast}=(y_{t}^{+},y_{t}^{-},x_{t}^{\trans})^{\trans}$
can be written as a linear function of elements of $(y_{t}^{+},x_{t}^{\trans})^{\trans}$
and $(y_{t}^{-},x_{t}^{\trans})^{\trans}$, it follows from \lemref{pmavg}
and the continuous mapping theorem, under the conditions of \lemref{pmavg}
and \assref{det} that
\begin{equation}
\frac{1}{n}\sum_{t=1}^{\smlfloor{n\lambda}}\left(g(w_{t})\otimes\begin{bmatrix}1\\
n^{-1/2}z_{t}^{\ast}
\end{bmatrix}\right)\indic^{\pm}(y_{t})\wkc[\expect g(w_{0}^{\pm})]\otimes\int_{0}^{\lambda}\begin{bmatrix}1\\
Z^{\ast}(\mu)
\end{bmatrix}\indic^{\pm}[Y(\mu)]\diff\mu\label{eq:Zastavg}
\end{equation}
on $D[0,1]$, jointly with $U_{n}\wkc U$.
\begin{lem}
\label{lem:comp}Suppose \assref{dgp-canon}, \assref{co}, \assref{coclas}
and \assref{det} hold. Then
\begin{enumerate}
\item \label{enu:comp:nonsingular}for all $\ctr_{xy}^{\pm}\in\reals^{q-1}$,
the matrix $[\beta^{\ast},\ctr^{\ast}]$ is nonsingular;
\item \label{enu:comp:limits}on $D[0,1]$,
\begin{equation}
\frac{1}{n}\sum_{t=1}^{\smlfloor{n\lambda}}\begin{bmatrix}\dmn{\varrho}_{n,t}\\
\dmn{\xi}_{t}
\end{bmatrix}\wkc\begin{bmatrix}\int_{0}^{\lambda}\dmn R(s)\diff s\\
0
\end{bmatrix}\label{eq:component}
\end{equation}
where
\[
\dmn R(s)\defeq R(s)-\int_{0}^{1}R(\lambda)\diff\lambda;
\]
\item \label{enu:comp:second}there exist positive definite matrices $\Sigma_{\xi^{+}}$
and $\Sigma_{\xi^{-}}$ such that 
\[
\frac{1}{n}\sum_{t=1}^{n}\begin{bmatrix}\dmn{\varrho}_{n,t}\dmn{\varrho}_{n,t}^{\trans} & \dmn{\varrho}_{n,t}\dmn{\xi}_{t}^{\trans}\\
\dmn{\xi}_{t}\dmn{\varrho}_{n,t}^{\trans} & \dmn{\xi}_{t}\dmn{\xi}_{t}^{\trans}
\end{bmatrix}\wkc\begin{bmatrix}\int_{0}^{1}\dmn R(s)\dmn R(s)^{\trans}\diff s & 0\\
0 & \Sigma_{\xi^{+}}m_{Y}^{+}(1)+\Sigma_{\xi^{-}}m_{Y}^{-}(1)
\end{bmatrix}
\]
and the r.h.s.\ is positive definite a.s.
\end{enumerate}
\end{lem}

Recall the definition of the $q$-dimensional standard (up to initialisation)
Brownian motion $W_{0}$ given in \eqref{stdbm}.
\begin{lem}
\label{lem:stdised} Suppose \assref{dgp-canon}, \assref{co}, \assref{coclas}
and \assref{det} hold. Then there exist $\ctr_{xy}^{\pm}\in\reals^{q-1}$,
and an invertible $(q+1)\times(q+1)$ matrix $Q$ such that
\[
QR(\lambda)=S_{q}[e_{1}^{\trans}W_{0}(\lambda)]W_{0}(\lambda)=W_{0}^{\ast}(\lambda).
\]
Moreover, there exist $\omega^{\pm}>0$ such that
\begin{equation}
Y(\lambda)=\omega^{+}[e_{1}^{\trans}W_{0}(\lambda)]_{+}+\omega^{-}[e_{1}^{\trans}W_{0}(\lambda)]_{-}.\label{eq:Yrep}
\end{equation}
\end{lem}

We note further that because the mapping between $z_{t}=(y_{t},x_{t}^{\trans})^{\trans}$
and its derived canonical form $\tilde{z}_{t}=(\tilde{y}_{t},\tilde{x}_{t}^{\trans})^{\trans}$
is such that $\tilde{y}_{t}^{+}$ and $\tilde{y}_{t}^{-}$ are respectively
positive scalar multiples of $y_{t}^{+}$ and $y_{t}^{-}$, a representation
of the form \eqref{Yrep} also obtains when \assref{dgp} holds in
place of \assref{dgp-canon}.
\begin{lem}
\label{lem:rank} Suppose $\mathcal{W}_{0}=0$ in \eqref{stdbm}.
Then the matrices
\begin{align*}
\dmn S_{W}^{\ast} & \defeq\int_{0}^{1}\dmn W_{0}^{\ast}(s)\dmn W_{0}^{\ast}(s)^{\trans}\diff s, & \dmn S_{V}^{\ast}\defeq & \int_{0}^{1}\dmn V_{0}^{\ast}(s)\dmn V_{0}^{\ast}(s)^{\trans}\diff s,
\end{align*}
are positive definite a.s.
\end{lem}

\section{Proofs of main results}

\label{app:thmproof}

\subsection{Proof of \propref{breitung}}

Since $\mathbb{A}_{n}$ and $\mathbb{B}_{n}$ are positive definite
with probability approaching one (w.p.a.1.), the eigenvalues $\{\lambda_{n,i}\}_{i=1}^{d_{w}}$
of $\mathbb{A}_{n}\mathbb{B}_{n}^{-1}$ are well defined, real and
positive w.p.a.1. By our assumptions and the continuous mapping theorem
(CMT),
\[
n^{-1}\mathbb{A}_{n}=\frac{1}{n}\sum_{t=1}^{n}w_{n,t}w_{n,t}^{\trans}\wkc\begin{bmatrix}\int_{0}^{1}\W(s)\W(s)^{\trans}\diff s & 0\\
0 & \Omega
\end{bmatrix},
\]
and 
\[
n^{-3}\mathbb{B}_{n}=\frac{1}{n}\sum_{t=1}^{n}\left(\frac{1}{n}\sum_{i=1}^{t}w_{n,i}\right)\left(\frac{1}{n}\sum_{j=1}^{t}w_{n,j}\right)^{\trans}\wkc\begin{bmatrix}\int_{0}^{1}\V(s)\V(s)^{\trans}\diff s & 0\\
0 & 0
\end{bmatrix}.
\]
Let $\{\mu_{n,i}\}_{i=1}^{d_{w}}$ denote the eigenvalues of $\mathbb{B}_{n}\mathbb{A}_{n}^{-1}$
ordered as $\mu_{n,1}\leq\mu_{n,2}\leq\cdots\leq\mu_{n,d_{w}}$, so
that $\lambda_{n,i}=\mu_{n,d_{w}+1-i}^{-1}$ for $1\le i\le d_{w}$.
By the CMT and the a.s.\ invertibility of $\Omega$,
\begin{align}
n^{-2}\mathbb{B}_{n}\mathbb{A}_{n}^{-1}=(n^{-3}\mathbb{B}_{n})(n^{-1}\mathbb{A}_{n})^{-1} & \wkc\begin{bmatrix}\int_{0}^{1}\V(s)\V(s)^{\trans}\diff s & 0\\
0 & 0
\end{bmatrix}\begin{bmatrix}\int_{0}^{1}\W(s)\W(s)^{\trans}\diff s & 0\\
0 & \Omega
\end{bmatrix}^{-1}\nonumber \\
 & =\begin{bmatrix}\int_{0}^{1}\V(s)\V(s)^{\trans}\diff s\left(\int_{0}^{1}\W(s)\W(s)^{\trans}\diff s\right)^{-1} & 0\\
0 & 0
\end{bmatrix}.\label{eq:BAcvg}
\end{align}
For the above limiting matrix, let $\{\mu_{i}^{\ast}\}_{i=1}^{d_{w}}$
denote its eigenvalues ordered as $\mu_{1}^{\ast}\leq\mu_{2}^{\ast}\leq\cdots\leq\mu_{d_{w}}^{\ast}$.
The first $d_{w}-\ell$ eigenvalues are zero, i.e.\ $\mu_{i}^{\ast}=0$
for $1\le i\le d_{w}-\ell$. The remaining $\ell$ eigenvalues $\{\mu_{i}^{\ast}\}_{i=d_{w}-\ell+1}^{d_{w}}$
are real and positive since they are the eigenvalues of
\[
\int_{0}^{1}\V(s)\V(s)^{\trans}\diff s\left(\int_{0}^{1}\W(s)\W(s)^{\trans}\diff s\right)^{-1}\eqdef{\cal V}{\cal W}^{-1},
\]
where ${\cal W}$ and ${\cal V}$ are positive definite almost surely.
By \eqref{BAcvg}, the continuity of eigenvalues and the CMT, then:
\begin{enumerate}
\item for $1\le i\le\ell$, 
\[
n^{2}\lambda_{n,i}=(n^{-2}\mu_{n,d_{w}+1-i})^{-1}\wkc(\mu_{d_{w}+1-i}^{\ast})^{-1}=(\mu_{(d_{w}-\ell)+(\ell+1-i)}^{\ast})^{-1}<\infty,
\]
where $\mu_{(d_{w}-\ell)+(\ell+1-i)}^{\ast}>0$ is the $(\ell+1-i)$th
eigenvalue of ${\cal V}{\cal W}^{-1}$; and
\item for $\ell+1\le i\le d_{w}$,
\[
n^{2}\lambda_{n,i}=(n^{-2}\mu_{n,d_{w}+1-i})^{-1}\inprob\infty,
\]
since $n^{-2}\mu_{n,d_{w}+1-i}\inprob\mu_{d_{w}+1-i}^{\ast}=0$.
\end{enumerate}
Therefore, if $\ell_{0}=\ell$ 
\begin{align*}
n^{2}\sum_{i=1}^{\ell_{0}}\lambda_{n,i}=\sum_{i=1}^{\ell_{0}}(n^{-2}\mu_{n,d_{w}+1-i})^{-1} & \wkc\sum_{i=1}^{\ell_{0}}(\mu_{(d_{w}-\ell)+(\ell+1-i)}^{\ast})^{-1}=\tr[({\cal V}{\cal W}^{-1})^{-1}]=\tr({\cal W}{\cal V}^{-1}),
\end{align*}
where the penultimate equality holds since the trace of a matrix equals
the sum of its eigenvalues; and if $\ell_{0}>\ell$,
\[
n^{2}\sum_{i=1}^{\ell_{0}}\lambda_{n,i}=n^{2}\sum_{i=1}^{\ell}\lambda_{n,i}+n^{2}\sum_{i=\ell+1}^{\ell_{0}}\lambda_{n,i}=n^{2}\sum_{i=1}^{\ell}\lambda_{n,i}+\sum_{i=\ell+1}^{\ell_{0}}(n^{-2}\mu_{n,d_{w}+1-i})^{-1}\inprob\infty,
\]
since $n^{2}\sum_{i=1}^{\ell}\lambda_{n,i}=O_{p}(1)$ and the second
term diverges in probability.\hfill\qedsymbol

\subsection{Proof of \thmref{lln}}

As noted in the proof of Theorem~4.4 in \DMW{}, the process $\{\tilde{z}_{t}\}$
obtained via the mapping \eqref{canon-vars} satisfies both \assref{dgp-canon},
and \ref{ass:detprime}. Thus $\{\tilde{z}_{t}\}$ satisfies the requirements
of \lemref{pmavg}. The convergence \eqref{gavg} follows immediately,
since $\sgn\tilde{y}_{t}=\sgn y_{t}$ by Proposition~2.1 of \citet{DMW23stat}.

We next proceed to establish the convergence \eqref{gzavg} holds
in the `$+$' case; the proof in the `$-$' case is analogous.
As per (D.3) of \DMW{}, define
\[
P(\pm1)\defeq\begin{bmatrix}\bar{\phi}_{0,yy}^{\pm} & 0\\
\phi_{0,xy}^{\pm} & \Phi_{0,xx}
\end{bmatrix}^{-1}
\]
and set $P(y)=P(+1)\indic^{+}(y)+P(-1)\indic^{-}(y)$. It follows
from (D.18) in \DMW{} that 
\[
z_{t}^{d}=P(\tilde{y}_{t})\tilde{z}_{t}^{d},
\]
and therefore
\[
\indic^{+}(y_{t})z_{t}^{d}=\indic^{+}(\tilde{y}_{t})P(\tilde{y}_{t})\tilde{z}_{t}^{d}=\indic^{+}(\tilde{y}_{t})P(+1)\tilde{z}_{t}^{d}.
\]
Since \eqref{gzavg} obtains for $\{\tilde{z}_{t}\}$ by \lemref{pmavg},
it follows that 
\begin{align*}
\frac{1}{n^{3/2}}\sum_{t=1}^{\smlfloor{n\lambda}}[g(w_{t})\otimes z_{t}^{d}]\indic^{+}(y_{t}) & =[I_{d_{g}}\otimes P(+1)]\frac{1}{n^{3/2}}\sum_{t=1}^{\smlfloor{n\lambda}}[g(w_{t})\otimes\tilde{z}_{t}^{d}]\indic^{+}(\tilde{y}_{t})\\
 & \wkc[I_{d_{g}}\otimes P(+1)][\expect g(w_{0}^{+})]\otimes\int_{0}^{\lambda}\tilde{Z}(\mu)\indic^{+}[\tilde{Y}(\mu)]\diff\mu\\
 & =[\expect g(w_{0}^{+})]\otimes\int_{0}^{\lambda}Z(\mu)\indic^{+}[Y(\mu)]\diff\mu,
\end{align*}
where we have used that
\begin{align*}
\indic^{+}[\tilde{Y}(\mu)]P(+1)\tilde{Z}(\mu) & =\indic^{+}[\tilde{Y}(\mu)]P[\tilde{Y}(\mu)]\tilde{Z}(\mu)\\
 & =\indic^{+}[Y(\mu)]Z(\mu)
\end{align*}
 as per (D.13) of \DMW{}.\hfill\qedsymbol

\subsection{Proof of \thmref{breitung}}

We now seek to verify the conditions of \propref{breitung}. As discussed
in \subsecref{canonical}, by Proposition~2.1 in \citet{DMW23stat}
there exists an invertible $P\in\reals^{(p+1)\times(p+1)}$ such that
\begin{equation}
\tilde{z}_{t}^{\ast}=\begin{bmatrix}\tilde{y}_{t}^{+}\\
\tilde{y}_{t}^{-}\\
\tilde{x}_{t}
\end{bmatrix}\defeq P^{-1}\begin{bmatrix}y_{t}^{+}\\
y_{t}^{-}\\
x_{t}
\end{bmatrix}=P^{-1}z_{t}^{\ast},\label{eq:canonP}
\end{equation}
where $\sgn\tilde{y}_{t}=\sgn y_{t}$. As noted in Remark~4.2(i)
of \DMW{}, $\{\tilde{z}_{t}\}$ follows -- in view of our assumptions,
in particular of the form taken by \assref{coclas}\ref{enu:coclas:jsr}
-- a canonical CKSVAR satisfying \assref{dgp-canon}, \assref{co},
\assref{coclas} and \assref{det}. Because of the invariance properties
of generalised eigenvalues, $\Lambda_{n,q_{0}}$ is invariant to the
pre- and/or post-multiplication of $\mathbf{A}_{n}$ and $\mathbf{B}_{n}$
by common matrices, and so it follows from \eqref{canonP} that $\Lambda_{n,q_{0}}$
computed on $\{z_{t}^{\ast}\}$ is identical to that computed on $\{\tilde{z}_{t}^{\ast}\}$.
We may therefore suppose, without loss of generality, that $\{z_{t}\}$
follows a canonical CKSVAR, i.e.\ that \assref{dgp-canon} holds
in place of \assref{dgp}.

By those same invariance properties of generalised eigenvalues, we
may further replace $(\mathbf{A}_{n},\mathbf{B}_{n})$ by 
\begin{align*}
\mathbb{A}_{n} & \defeq\bar{Q}(\breit_{n}^{\trans}\mathbf{A}_{n}\breit_{n})\bar{Q}^{\trans}=\sum_{t=1}^{n}w_{n,t}w_{n,t}^{\trans} & \mathbb{B}_{n} & \defeq\bar{Q}(\breit_{n}^{\trans}\mathbf{B}_{n}\breit_{n})\bar{Q}^{\trans}=\sum_{t=1}^{n}\sum_{i=1}^{t}w_{n,i}\sum_{j=1}^{t}w_{n,j}^{\trans}
\end{align*}
where where $\bar{Q}\defeq\diag\{Q,I_{r}\}$, for $Q$ as in \lemref{stdised},
and as per \eqref{demeaned},
\[
w_{n,t}\defeq\bar{Q}(\breit_{n}^{\trans}\dmn z_{t}^{\ast})=\begin{bmatrix}Q\dmn{\varrho}_{n,t}\\
\dmn{\xi}_{t}
\end{bmatrix}.
\]
By Lemmas~\ref{lem:comp} and \ref{lem:stdised}, $\{w_{n,t}\}$
satisfies the requirements of \propref{breitung}, with
\begin{align}
\mathbb{W}(s) & =Q\dmn R(s)=\bar{W}_{0}^{\ast}(s), & \Omega & =\Sigma_{\xi^{+}}m_{Y}^{+}(1)+\Sigma_{\xi^{-}}m_{Y}^{-}(1),\label{eq:breitunglim}
\end{align}
and $\mathbb{V}(s)=\dmn V_{0}^{\ast}(s)=\int_{0}^{s}\dmn W_{0}^{\ast}(\lambda)\diff\lambda$,
with the a.s.\ positive definiteness of $\int_{0}^{1}\W(s)\W(s)^{\trans}\diff s$
and $\int_{0}^{1}\V(s)\V(s)^{\trans}\diff s$ following by \lemref{rank}.

An application of \propref{breitung} (with $\ell=q+1$ and $\ell_{0}=q_{0}+1$)
then yields the conclusions of parts~\ref{enu:qeqq0} and \ref{enu:q0gtq}.
Part~\ref{enu:q0ltq} follows immediately from the result of part~\ref{enu:qeqq0},
noting that $\Lambda_{n,q_{0}}\leq\Lambda_{n,q}$ for all $n$, in
this case, and $\Lambda_{n,q}\wkc\Lambda_{q}$. Under \assref{dgp-canon},
the convergence in \eqref{YW0} is an immediate consequence of \lemref{stdised};
if instead \assref{dgp} holds, then this follows from the fact that
$y_{t}^{+}$ and $y_{t}^{-}$ are respectively scalar multiples of
the canonical variables $\tilde{y}_{t}^{+}$ and $\tilde{y}_{t}^{-}$,
by \eqref{canon-vars}.\hfill\qedsymbol

\section{Proofs of auxiliary lemmas}
\begin{proof}[Proof of \lemref{YZwkc}]
\textbf{\ref{enu:YZwkc:ydelta}.} We have
\[
\frac{1}{n}\sum_{t=1}^{n}\indic\{n^{-1/2}\smlabs{y_{t}}\leq\delta\}=\frac{1}{n}\sum_{t=1}^{n}\indic\{-\delta\leq n^{-1/2}y_{t}<0\}+\frac{1}{n}\sum_{t=1}^{n}\indic\{0\leq n^{-1/2}y_{t}\leq\delta\}.
\]
We will show that the second r.h.s.\ term is $o_{p}(1)$ as $n\goesto\infty$
and then $\delta\goesto0$; the proof for the first r.h.s.\ term
is analogous. Similarly to the proof of Theorem~4.2 in \DMW{}, define
$f(y)\defeq h(y)^{-1}y$. Then $f(y)=y$ for all $y\geq0$, and it
follows from \eqref{Zncvg} and \eqref{Ylim} above that
\[
f(n^{-1/2}y_{\smlfloor{n\lambda}})\wkc f[Y(\lambda)]=\vartheta^{\trans}U_{0}(\lambda)=\vartheta^{\trans}\Gamma(1;\mathcal{Y}_{0})\mathcal{Z}_{0}+\vartheta^{\trans}U(\lambda)\eqdef{\cal B}_{0}+B(\lambda)
\]
where $B$ is a (scalar) Brownian motion, and ${\cal B}_{0}\in\reals$
is non-random. Since $x\elmap\indic\{0<x\leq\delta\}$ is Riemann
integrable, it follows\textcolor{red}{{} }\textcolor{black}{by Theorem~2.3
and Remark~2.2 in \citet{BH06ET}} that
\begin{align*}
\frac{1}{n}\sum_{t=1}^{n}\indic\{0\leq n^{-1/2}y_{t}\leq\delta\} & =\frac{1}{n}\sum_{t=1}^{n}\indic\{0\leq f(n^{-1/2}y_{t})\leq\delta\}\\
 & \wkc\int_{0}^{1}\indic\{0\leq{\cal B}_{0}+B(\lambda)\leq\delta\}\diff\lambda\inprob0
\end{align*}
as $n\goesto\infty$ and then $\delta\goesto0$, since $B$ has a
(Lebesgue) local time density.

\textbf{\ref{enu:YZwkc:YZ}.} By the Cramér--Wold device, it suffices
to show that, on $D[0,1]$,
\[
\frac{1}{n}\sum_{t=1}^{\smlfloor{n\lambda}}\indic^{\pm}(y_{t})(a_{0}+a^{\trans}z_{n,t}^{d})\wkc\int_{0}^{\lambda}\indic^{\pm}[Y(\mu)][a_{0}+a^{\trans}Z(\mu)]\diff\mu
\]
for $a_{0}\in\reals$ and $a\in\reals^{p}$, where $z_{n,t}^{d}\defeq n^{-1/2}z_{t}^{d}$.
We give the proof here for $\indic^{+}$; the proof for $\indic^{-}$
is analogous. To that end, define
\[
T(\lambda)\defeq\int_{0}^{\lambda}\indic^{+}[Y(\mu)][a_{0}+a^{\trans}Z(\mu)]\diff\mu.
\]

Letting $y_{n,t}\defeq n^{-1/2}y_{t}$, we have
\[
\frac{1}{n}\sum_{t=1}^{\smlfloor{n\lambda}}\indic^{+}(y_{t})(a_{0}+a^{\trans}z_{n,t}^{d})=\frac{1}{n}\sum_{t=1}^{\smlfloor{n\lambda}}\indic\{y_{n,t}\geq0\}(a_{0}+a^{\trans}z_{n,t}^{d})\eqdef T_{n}(\lambda)
\]
For $\epsilon>0$, define a continuous function
\[
f_{\epsilon}(y)\defeq\begin{cases}
0 & \text{if }y<0\\
\frac{1}{\epsilon}y & \text{if }y\in[0,\epsilon)\\
1 & \text{if }y\geq\epsilon,
\end{cases}
\]
so that by CMT and \eqref{zd-cvg},
\[
T_{n,\epsilon}(\lambda)\defeq\frac{1}{n}\sum_{t=1}^{\smlfloor{n\lambda}}f_{\epsilon}(y_{n,t})(a_{0}+a^{\trans}z_{n,t}^{d})\wkc\int_{0}^{\lambda}f_{\epsilon}[Y(\mu)][a_{0}+a^{\trans}Z(\mu)]\diff\mu\eqdef T_{\epsilon}(\lambda)
\]
as $n\goesto\infty$. It then follows by arguments given in the proof
of part~\ref{enu:YZwkc:ydelta} that, for some $C<\infty$ (depending
on $a$ and $a_{0}$),
\begin{align*}
\smlabs{T_{\epsilon}(\lambda)-T(\lambda)} & \leq C\left(1+\sup_{\lambda\in[0,1]}\smlnorm{Z(\lambda)}\right)\int_{0}^{1}\indic\{0\leq Y(\mu)\leq\epsilon\}\diff\mu\\
 & =C\left(1+\sup_{\lambda\in[0,1]}\smlnorm{Z(\lambda)}\right)\int_{0}^{1}\indic\{0\leq\mathcal{B}_{0}+B(\mu)\leq\epsilon\}\diff\mu\inprob0
\end{align*}
as $\epsilon\goesto0$. Moreover, by the result of part~\ref{enu:YZwkc:ydelta},
and \eqref{supbnd},
\begin{align*}
\smlabs{T_{n,\epsilon}(\lambda)-T_{n}(\lambda)} & \leq\frac{1}{n}\sum_{t=1}^{\smlfloor{n\lambda}}\smlabs{f_{\epsilon}(y_{n,t})-\indic^{+}(y_{n,t})}\smlabs{a_{0}+a^{\trans}z_{n,t}^{d}}\\
 & \leq C\left(1+\sup_{1\leq s\leq n}\smlnorm{z_{n,s}^{d}}\right)\frac{1}{n}\sum_{t=1}^{n}\indic\{0\leq y_{n,t}\leq\epsilon\}\inprob0
\end{align*}
as $n\goesto\infty$ and then $\epsilon\goesto0$. The preceding three
convergences thus yield the result.
\end{proof}

\begin{proof}[Proof of \lemref{pmavg}]
 By the Cramér-Wold device, it suffices to consider the case where
$d_{g}=1$. We note that the r.h.s.\ of \eqref{w0} is well defined
since $\rho(A^{\pm})\leq\rho_{\jsr}({\cal A})<1$. Here we shall prove
the results only in the `$+$' case; the proof in the `$-$' case
follows by identical arguments. We also only give the proof of \eqref{gzavg},
since \eqref{gavg} is essentially a simpler case of \eqref{gzavg}
in which $n^{-1/2}z_{t}^{d}\eqdef z_{n,t}^{d}$ has been replaced
by $1$. The proof proceeds in the following five steps.
\begin{enumerate}
\item Reduction to the case where $g$ is bounded.
\item Disentangling of weakly dependent and integrated components:
\begin{align}
\frac{1}{n}\sum_{t=1}^{\smlfloor{n\lambda}}g(w_{t})z_{n,t}^{d}\indic^{+}(y_{t}) & =\frac{1}{n}\sum_{t=1}^{\smlfloor{n\lambda}}g(w_{t})z_{n,t-m}^{d}\indic\{y_{t-m}\geq n^{1/2}\delta\}+o_{p}(1)\label{eq:disentangle}
\end{align}
as $n\goesto\infty$, $m\goesto\infty$ and then $\delta\goesto0$,
uniformly over $\lambda\in[0,1]$.
\item Approximation of $w_{t}$: for each $m\in\naturals$ and $\delta>0$,
\begin{equation}
\frac{1}{n}\sum_{t=1}^{\smlfloor{n\lambda}}g(w_{t})z_{n,t-m}^{d}\indic\{y_{t-m}\geq n^{1/2}\delta\}=\frac{1}{n}\sum_{t=1}^{\smlfloor{n\lambda}}g(w_{m,t}^{+})z_{n,t-m}^{d}\indic\{y_{t-m}\geq n^{1/2}\delta\}+o_{p}(1)\label{eq:wapprox}
\end{equation}
as $n\goesto\infty$, uniformly over $\lambda\in[0,1]$, where 
\begin{equation}
w_{m,t}^{+}\defeq\sum_{\ell=0}^{m-1}(A^{+})^{\ell}(c^{+}+B^{+}v_{t-\ell}).\label{eq:wmtp}
\end{equation}
\item Recentring of $g(w_{m,t}^{+})$: for each $m\in\naturals$ and $\delta>0$,
\[
\frac{1}{n}\sum_{t=1}^{\smlfloor{n\lambda}}g(w_{m,t}^{+})z_{n,t-m}^{d}\indic\{y_{t-m}\geq n^{1/2}\delta\}=[\expect g(w_{m,0}^{+})]\frac{1}{n}\sum_{t=1}^{\smlfloor{n\lambda}}z_{n,t-m}^{d}\indic\{y_{t-m}\geq n^{1/2}\delta\}+o_{p}(1)
\]
as $n\goesto\infty$, uniformly over $\lambda\in[0,1]$.
\item Computing the limit:
\[
[\expect g(w_{m,0}^{+})]\frac{1}{n}\sum_{t=1}^{\smlfloor{n\lambda}}z_{n,t-m}^{d}\indic\{y_{t-m}\geq n^{1/2}\delta\}\wkc[\expect g(w_{0}^{+})]\int_{0}^{\lambda}Z(\mu)\indic^{+}[Y(\mu)]\diff\mu
\]
on $D[0,1]$, as $n\goesto\infty$, $m\goesto\infty$ and then $\delta\goesto0$.
\end{enumerate}

\subparagraph{(i) Reduction to the case where $g$ is bounded.}

It follows directly from the local Lipschitz condition on $g$ that
\begin{equation}
\smlabs{g(w)}\leq\smlabs{g(0)}+C(1+\smlnorm w^{\ell_{0}})\smlnorm w\leq C_{1}(1+\smlnorm w^{\ell_{0}+1})\label{eq:gbound}
\end{equation}
for all $w\in\reals^{d_{w}}$, and hence for some $\eta_{0}\in(0,m_{0}/(\ell_{0}+1)-1]$,
which exists since $m_{0}>\ell_{0}+1$,
\[
\smlabs{g(w)}^{1+\eta_{0}}\leq C_{2}(1+\smlnorm w^{(\ell_{0}+1)(1+\eta_{0})})\leq C_{3}(1+\smlnorm w^{m_{0}}).
\]
Since $\sup_{t\in\integers}\smlnorm{w_{t}}_{m_{0}}<\infty$ by Lemma~A.1
in \DMW{}, it follows immediately that $\sup_{t\in\integers}\smlnorm{g(w_{t})}_{1+\eta_{0}}<\infty$.
Moreover, since
\begin{equation}
\smlnorm{w_{0}^{+}}_{m_{0}}\leq\smlnorm{(I_{d_{w}}-A^{+})^{-1}c^{+}}+\sum_{\ell=0}^{\infty}\smlnorm{(A^{+})^{\ell}}\smlnorm{B^{+}}\smlnorm{v_{-\ell}}_{m_{0}}<\infty,\label{eq:w0mom}
\end{equation}
it follows that $\expect|g(w_{0}^{+})|^{1+\eta_{0}}<\infty$, so that
the r.h.s.\ of \eqref{gzavg} is indeed well defined.

Now decompose
\[
g(w)=g(w)\indic\{\smlabs{g(w)}\leq M\}+g(w)\indic\{\smlabs{g(w)}>M\}\eqdef g_{M}^{(\leq)}(w)+g_{M}^{(>)}(w).
\]
Recalling $z_{n,t}^{d}\defeq n^{-1/2}z_{t}^{d}$, we have
\[
\abs{\frac{1}{n}\sum_{t=1}^{\smlfloor{n\lambda}}g_{M}^{(>)}(w_{t})z_{n,t}^{d}\indic^{+}(y_{t})}\leq\sup_{s\leq n}\smlnorm{z_{n,s}^{d}}\frac{1}{n}\sum_{t=1}^{n}\smlabs{g_{M}^{(>)}(w_{t})}\inprob0
\]
as $n\goesto\infty$ and then $M\goesto\infty$, since $\sup_{s\leq n}\smlnorm{z_{n,s}^{d}}=O_{p}(1)$
as per \eqref{supbnd} above, and by Chebyshev's inequality,
\[
\sup_{t\in\integers}\expect\smlabs{g_{M}^{(>)}(w_{t})}\leq\frac{\sup_{t\in\integers}\expect\smlabs{g(w_{t})}^{1+\eta_{0}}}{M^{\eta_{0}}}\goesto0
\]
as $M\goesto\infty$. Since $\expect g_{M}^{(>)}(w_{0}^{+})\goesto0$
as $M\goesto\infty$ by dominated convergence, it suffices to prove
the result with $g_{M}^{(\leq)}$ in place of $g$. Moreover, since
$g_{M}^{(\leq)}$ satisfies the same local Lipschitz condition as
does $g$, we may henceforth suppose that $g$ itself is bounded by
some constant $C_{g}<\infty$, without loss of generality.

\subparagraph{(ii) Disentangling of weakly dependent and integrated components.}

Let $m\in\naturals$. Since $d_{g}=1$, we have that $g(w_{t})\otimes z_{t}^{d}=g(w_{t})z_{t}^{d}$.
The l.h.s.\ of \eqref{gzavg} may be written as
\begin{equation}
\frac{1}{n}\sum_{t=1}^{\smlfloor{n\lambda}}g(w_{t})z_{n,t}^{d}\indic^{+}(y_{t})=\sum_{i=0}^{m-1}\frac{1}{n}\sum_{t=1}^{\smlfloor{n\lambda}}g(w_{t})\Delta z_{n,t-i}^{d}\indic^{+}(y_{t})+\frac{1}{n}\sum_{t=1}^{\smlfloor{n\lambda}}g(w_{t})z_{n,t-m}^{d}\indic^{+}(y_{t}),\label{eq:gdecomp}
\end{equation}
where we recall the convention that $\Delta z_{i}=\Delta z_{i}^{d}=0$
for all $i\leq-k$ and that therefore $z_{i}^{d}=z_{i}=z_{-k}=z_{-k}^{d}$
for all $i\leq-k$, as per \eqref{initdiff} above. For each $i\in\{0,\ldots,m-1\}$,
we have 
\begin{equation}
\norm{\frac{1}{n}\sum_{t=1}^{\smlfloor{n\lambda}}g(w_{t})\Delta z_{n,t-i}^{d}\indic^{+}(y_{t})}\leq C_{g}\sup_{s\leq n}\smlnorm{\Delta z_{n,s}^{d}}\inprob0\label{eq:negl1}
\end{equation}
as $n\goesto\infty$, since $\sup_{s\leq n}\smlnorm{\Delta z_{n,s}^{d}}=o_{p}(1)$
by \eqref{supbnd}. Deduce that the first r.h.s.\ term in \eqref{gdecomp}
is $o_{p}(1)$ as $n\goesto\infty$, uniformly in $\lambda\in[0,1]$.

This leaves the second r.h.s.\ term in \eqref{gdecomp}; to complete
the proof of \eqref{disentangle}, we need to replace $\indic^{+}(y_{t})=\indic\{y_{t}\geq0\}$
by $\indic\{y_{t-m}\geq n^{1/2}\delta\}$. Therefore consider
\begin{align*}
\smlabs{\indic\{y_{t-m}\geq n^{1/2}\delta\}-\indic^{+}(y_{t})} & =\indic\{y_{t}\leq0\sep y_{t-m}\geq n^{1/2}\delta\}+\indic\{y_{t}\geq0\sep y_{t-m}\leq n^{1/2}\delta\}\\
 & \leq\indic\{y_{t}\leq0\sep y_{t-m}\geq n^{1/2}\delta\}\\
 & \qquad\qquad+\indic\{y_{t}\geq0\sep y_{t-m}\leq-n^{1/2}\delta\}+\indic\{\smlabs{y_{t-m}}<n^{1/2}\delta\}\\
 & \eqdef\kappa_{1t}+\kappa_{2t}+\kappa_{3t}
\end{align*}
Using that $y_{t}-y_{t-m}=\sum_{\ell=0}^{m-1}\Delta y_{t-\ell}$,
we have
\begin{equation}
y_{t}\leq0\text{ and }y_{t-m}\geq n^{1/2}\delta\implies\abs{\sum_{\ell=0}^{m-1}\Delta y_{t-\ell}}\geq n^{1/2}\delta.\label{eq:signdiff}
\end{equation}
Hence
\begin{align*}
\frac{1}{n}\sum_{t=1}^{n}\kappa_{1t} & \leq\frac{1}{n}\sum_{t=1}^{n}\indic\left\{ \abs{\sum_{\ell=0}^{m-1}\Delta y_{t-\ell}}\geq n^{1/2}\delta\right\} \\
 & \leq\sum_{\ell=0}^{m-1}\frac{1}{n}\sum_{t=1}^{n}\indic\{\smlabs{\Delta y_{t-\ell}}\geq n^{1/2}m^{-1}\delta\}
\end{align*}
where the second inequality holds since if $am\leq\smlabs{\sum_{\ell=0}^{m-1}\Delta y_{t-\ell}}\leq\sum_{\ell=0}^{m-1}\smlabs{\Delta y_{t-\ell}}$,
then $\smlabs{\Delta y_{t-\ell}}\geq a$ for some $\ell\in\{0,\ldots,m-1\}$.
By Chebyshev's inequality,
\begin{equation}
\max_{t\leq n}\Prob\{\smlabs{\Delta y_{t}}\geq n^{1/2}m^{-1}\delta\}\leq n^{-1/2}\delta^{-1}m\max_{t\leq n}\expect\smlabs{\Delta y_{t}}\goesto0\label{eq:deltaycheby}
\end{equation}
as $n\goesto\infty$, since $\max_{t\leq n}\expect\smlabs{\Delta y_{t}}<\infty$
in view of \eqref{dzdmom}. Deduce that
\begin{equation}
\norm{\frac{1}{n}\sum_{t=1}^{\smlfloor{n\lambda}}g(w_{t})z_{n,t-m}^{d}\kappa_{1t}}\leq C_{g}\sup_{s\leq n}\smlnorm{z_{n,s}^{d}}\frac{1}{n}\sum_{t=1}^{n}\kappa_{1t}\inprob0.\label{eq:kap1}
\end{equation}
By a symmetric argument, the preceding also holds with $\kappa_{2t}$
in place of $\kappa_{1t}$. Finally, it follows from \lemref{YZwkc}\ref{enu:YZwkc:ydelta}
that
\begin{equation}
\norm{\frac{1}{n}\sum_{t=1}^{\smlfloor{n\lambda}}g(w_{t})z_{n,t-m}^{d}\kappa_{3t}}\leq C_{g}\sup_{s\leq n}\smlnorm{z_{n,s}^{d}}\frac{1}{n}\sum_{t=1}^{n}\indic\{\smlabs{y_{t-m}}<n^{1/2}\delta\}\inprob0\label{eq:kap3}
\end{equation}
as $n\goesto\infty$ and then $\delta\goesto0$. Thus \eqref{disentangle}
follows from \eqref{kap1} and \eqref{kap3}.

\subparagraph{(iii) Approximation of $w_{t}$.}

We begin by decomposing
\[
g(w_{t})=g(w_{m,t}^{+})+[g(w_{t})-g(w_{m,t}^{+})]\eqdef g(w_{m,t}^{+})+\nabla_{m,t}.
\]
Since $g$ is bounded, and $\sup_{s\leq n}\smlnorm{z_{n,s}^{d}}=O_{p}(1)$
as per \eqref{supbnd} above, the first $m$ summands on the l.h.s.\ of
\eqref{wapprox} are $o_{p}(1)$. Thus to prove \eqref{wapprox},
it suffices to establish the asymptotic negligiblility of 
\[
\norm{\frac{1}{n}\sum_{t=m+1}^{\smlfloor{n\lambda}}\nabla_{m,t}z_{n,t-m}^{d}\indic\{y_{t-m}\geq n^{1/2}\delta\}}\leq\sup_{s\leq n}\smlnorm{z_{n,s}^{d}}\frac{1}{n}\sum_{t=m+1}^{n}\smlabs{\nabla_{m,t}}\indic\{y_{t-m}\geq n^{1/2}\delta\}.
\]
To handle the sum on the r.h.s., define
\[
\indic_{m,t}\defeq\{y_{s}>0\sep\forall s\in\{t-m,\ldots,t\}\}.
\]
If $\indic_{m,t}=1$, then $y_{s}>0$ for all $s\in\{t-m,\ldots,t\}$,
and so $(A_{s},B_{s},c_{s})=(A^{+},B^{+},c^{+})$ for all $s\in\{t-m+1,\ldots,t\}$,
whence recursive substitution applied to \eqref{wproc} yields
\[
w_{t}=(A^{+})^{m}w_{t-m}+\sum_{\ell=0}^{m-1}(A^{+})^{\ell}(c^{+}+B^{+}v_{t-\ell})=(A^{+})^{m}w_{t-m}+w_{m,t}^{+}.
\]
In other words, when $\indic_{m,t}=1$ holds $w_{t}$ may be approximated
by $w_{m,t}^{+}$, and so $\nabla_{m,t}$ should be small. Indeed,
\begin{align*}
\smlabs{\nabla_{m,t}}\indic_{m,t}=\smlabs{g(w_{t})-g(w_{m,t}^{+})}\indic_{m,t} & =\smlabs{g[(A^{+})^{m}w_{t-m}+w_{m,t}^{+}]-g(w_{m,t}^{+})}\indic_{m,t}\\
 & \leq C_{1}\min\{1,\smlnorm{(A^{+})^{m}}\smlnorm{w_{t-m}}(1+\smlnorm{w_{t}}^{\ell_{0}}+\smlnorm{w_{m,t}^{+}}^{\ell_{0}})\}\\
 & \leq C_{2}\min\{1,\smlnorm{(A^{+})^{m}}\smlnorm{w_{t-m}}(1+\smlnorm{w_{t-m}}^{\ell_{0}}+\smlnorm{w_{m,t}^{+}}^{\ell_{0}})\}
\end{align*}
for some $C_{1},C_{2}<\infty$, using the local Lipschitz condition
\eqref{lip}, and the boundedness of $g$. By Lemma~A.1 of \DMW{},
for $\gamma\in(\rho_{\jsr}({\cal A}),1)$,
\begin{align*}
\smlnorm{w_{t}} & \leq C_{3}\left[\sum_{s=0}^{t-1}\gamma^{s}(1+\smlnorm{v_{t-s}})+\gamma^{t}\smlnorm{w_{0}}\right]
\end{align*}
for some $C_{3}<\infty$. Therefore, for $t\geq m+1$, the distribution
of $\smlnorm{w_{t-m}}$ is stochastically dominated by that of
\[
C_{3}\left[\sum_{\ell=1}^{\infty}\gamma^{\ell-1}(1+\smlnorm{v_{\ell}})+\smlnorm{w_{0}}\right]\eqdef\bar{w}_{0}
\]
while the distribution of $\smlnorm{w_{m,t}^{+}}$ is stochastically
dominated by that of
\[
\sum_{\ell=0}^{\infty}\smlnorm{(A^{+})^{\ell}}(\smlnorm{c^{+}}+\smlnorm{B^{+}}\smlnorm{v_{\ell}})\eqdef\bar{w}_{0}^{+}
\]
Since $w_{m,t}^{+}$ depends only on $\{v_{s}\}_{s=t-m+1}^{t}$, it
is independent of $w_{t-m}$. Therefore, taking $(\tilde{w}_{0},\tilde{w}_{0}^{+})$
to be such that $\tilde{w}_{0}$ and $\tilde{w}_{0}^{+}$ are independent,
with (marginally) $\tilde{w}_{0}\eqdist\bar{w}_{0}$ and $\tilde{w}_{0}^{+}\eqdist\bar{w}_{0}^{+}$,
we have that
\begin{align*}
\max_{m+1\leq t\leq n}\expect\smlabs{\nabla_{m,t}}\indic_{m,t} & \leq\max_{m+1\leq t\leq n}C_{2}\expect\min\{1,\smlnorm{(A^{+})^{m}}\smlnorm{w_{t-m}}(1+\smlnorm{w_{t-m}}^{\ell_{0}}+\smlnorm{w_{m,t}^{+}}^{\ell_{0}})\}\\
 & \leq C_{2}\expect\min\{1,\smlnorm{(A^{+})^{m}}\smlnorm{\tilde{w}_{0}}(1+\smlnorm{\tilde{w}_{0}}^{\ell_{0}}+\smlnorm{\tilde{w}_{0}^{+}}^{\ell_{0}})\}\\
 & \goesto0
\end{align*}
as $m\goesto\infty$, by dominated convergence. Deduce
\begin{align}
\frac{1}{n}\sum_{t=m+1}^{n}\smlabs{\nabla_{m,t}}\indic\{y_{t-m}\geq n^{1/2}\delta\} & =\frac{1}{n}\sum_{t=m+1}^{n}\smlabs{\nabla_{m,t}}\indic\{y_{t-m}\geq n^{1/2}\delta\}[\indic_{m,t}+(1-\indic_{m,t})]\nonumber \\
 & =\frac{1}{n}\sum_{t=m+1}^{n}\smlabs{\nabla_{m,t}}\indic\{y_{t-m}\geq n^{1/2}\delta\}(1-\indic_{m,t})+o_{p}(1).\label{eq:nablabnd}
\end{align}
as $n\goesto\infty$ and then $m\goesto\infty$.

It remains to show that the first r.h.s.\ term in \eqref{nablabnd}
is also asymptotically negligible. We note that the summands are nonzero
only if $\indic_{m,t}=0$, in which case, there must exist an $i\in\{0,\ldots,m\}$
such that $y_{t-i}\leq0$. Using a similar argument to that which
follows \eqref{signdiff} above, since $y_{t-i}=y_{t-m}+\sum_{j=i}^{m-1}\Delta y_{t-j}$
we have that
\[
y_{t-i}\leq0\text{ and }y_{t-m}\geq n^{1/2}\delta\implies\abs{\sum_{j=i}^{m-1}\Delta y_{t-j}}\geq n^{1/2}\delta.
\]
Hence
\begin{align}
\frac{1}{n}\sum_{t=1}^{n}\indic\{y_{t-m}\geq n^{1/2}\delta\}(1-\indic_{m,t}) & =\frac{1}{n}\sum_{t=1}^{n}\indic\{y_{t-m}\geq n^{1/2}\delta\}\indic\{\exists i\in\{0,\ldots,m\}\text{ s.t. }y_{t-i}\leq0\}\nonumber \\
 & \leq\sum_{i=0}^{m-1}\frac{1}{n}\sum_{t=1}^{n}\indic\{y_{t-m}\geq n^{1/2}\delta\}\indic\{y_{t-i}\leq0\}\nonumber \\
 & \leq\sum_{i=0}^{m-1}\frac{1}{n}\sum_{t=1}^{n}\indic\left\{ \abs{\sum_{j=i}^{m-1}\Delta y_{t-j}}\geq n^{1/2}\delta\right\} \nonumber \\
 & \leq\sum_{i=0}^{m-1}\sum_{j=i}^{m-1}\frac{1}{n}\sum_{t=1}^{n}\indic\{\smlabs{\Delta y_{t-j}}\geq n^{1/2}(m-i)^{-1}\delta\}\label{eq:kappa2-1}
\end{align}
with the expectation of the summands being bounded by the l.h.s.\ of
\eqref{deltaycheby}, modulo the replacement of $m$ by $m-i$ there.
Since $g$ is bounded, deduce that
\[
\frac{1}{n}\sum_{t=1}^{n}\smlabs{\nabla_{m,t}}\indic\{y_{t-m}\geq n^{1/2}\delta\}(1-\indic_{m,t})\inprob0
\]
as $n\goesto\infty$, as required.

\subparagraph{(iv) Recentring of $g(w_{m,t}^{+})$.}

Defining 
\[
\bar{g}(w_{m,t}^{+})\defeq g(w_{m,t}^{+})-\expect g(w_{m,t}^{+})=g(w_{m,t}^{+})-\expect g(w_{m,0}^{+})
\]
we may write
\begin{multline}
\frac{1}{n}\sum_{t=1}^{\smlfloor{n\lambda}}g(w_{m,t}^{+})z_{n,t-m}^{d}\indic\{y_{t-m}\geq n^{1/2}\delta\}=[\expect g(w_{m,0}^{+})]\frac{1}{n}\sum_{t=1}^{\smlfloor{n\lambda}}z_{n,t-m}^{d}\indic\{y_{t-m}\geq n^{1/2}\delta\}\\
+\frac{1}{n}\sum_{t=1}^{\smlfloor{n\lambda}}\bar{g}(w_{m,t}^{+})z_{n,t-m}^{d}\indic\{y_{t-m}\geq n^{1/2}\delta\}.\label{eq:gfinal}
\end{multline}
We must show that the second r.h.s.\ term in \eqref{gfinal} is negligible.
We first note that
\[
\expect\norm{\frac{1}{n}\sum_{t=1}^{\smlfloor{n\lambda}}\bar{g}(w_{m,t}^{+})z_{n,t-m}^{d}\indic\{y_{t-m}\geq n^{1/2}\delta\}\indic\{\smlnorm{z_{n,t-m}^{d}}>M\}}\leq C\Prob\left\{ \sup_{1\leq t\leq n}\smlnorm{z_{n,t}^{d}}>M\right\} \goesto0
\]
as $n\goesto\infty$ and then $M\goesto\infty$, since $\sup_{1\leq t\leq n}\smlnorm{z_{n,t}^{d}}=O_{p}(1)$.
Therefore, letting $h_{M}(z)\defeq z\indic\{\smlnorm z\leq M\}$,
it suffices to show that
\[
\frac{1}{n}\sum_{t=1}^{\smlfloor{n\lambda}}\bar{g}(w_{m,t}^{+})h_{M}(z_{n,t-m}^{d})\indic\{y_{t-m}\geq n^{1/2}\delta\}\inprob0
\]
as $n\goesto\infty$, for each $M>0$.

In view of \eqref{wmtp}, $w_{m,t}^{+}$ is a function only of $\{v_{t-m+1},\ldots,v_{t}\}$,
and is therefore independent of $\filt_{t-m}$. $\bar{g}(w_{m,t}^{+})$
admits the telescoping sum decomposition
\begin{align*}
\bar{g}(w_{m,t}^{+}) & =g(w_{m,t}^{+})-\expect g(w_{m,t}^{+})=\sum_{\ell=0}^{m-1}[\expect_{t-\ell}g(w_{m,t}^{+})-\expect_{t-\ell-1}g(w_{m,t}^{+})]\eqdef\sum_{\ell=0}^{m-1}\varsigma_{\ell,m,t},
\end{align*}
where $\expect_{s}[\cdot]\defeq\expect[\cdot\mid\filt_{s}]$, and
we have used the fact that $\expect_{t-m}g(w_{m,t}^{+})=\expect g(w_{m,t}^{+})$.
For every $\ell\in\{0,\ldots,m-1\}$, $\{\varsigma_{\ell,m,t}\}_{t\in\naturals}$
defines a bounded martingale difference sequence. Rewriting
\begin{multline}
\frac{1}{n}\sum_{t=1}^{\smlfloor{n\lambda}}\bar{g}(w_{m,t}^{+})h_{M}(z_{n,t-m}^{d})\indic\{y_{t-m}\geq n^{1/2}\delta\}\\
=\frac{1}{n^{1/2}}\sum_{\ell=0}^{m-1}\sum_{t=1}^{\smlfloor{n\lambda}}\frac{\varsigma_{\ell,m,t}}{n^{1/2}}h_{M}(z_{n,t-m}^{d})\indic\{y_{t-m}\geq n^{1/2}\delta\}\eqdef\frac{1}{n^{1/2}}\sum_{\ell=0}^{m-1}S_{\ell,m,n}(\lambda).\label{eq:gbar}
\end{multline}
Applying Theorem~2.11 in \citet[with $p=2$]{HH80} to each element
of the martingale $S_{\ell,m,n}(\lambda)$, it follows that there
exists a $C<\infty$ such that
\[
\expect\sup_{\lambda\in[0,1]}\smlnorm{S_{\ell,m,n}(\lambda)}^{2}\leq C(1+n^{-1}M^{2}),
\]
and hence
\[
\frac{1}{n^{1/2}}\sum_{\ell=0}^{m-1}S_{\ell,m,n}(\lambda)\inprob0
\]
uniformly in $\lambda\in[0,1]$, as $n\goesto\infty$.

\subparagraph{(v) Computing the limit.}

Finally, regarding the first r.h.s.\ term in \eqref{gfinal}, we
have
\[
\frac{1}{n}\sum_{t=1}^{\smlfloor{n\lambda}}z_{n,t-m}^{d}\indic\{y_{t-m}\geq n^{1/2}\delta\}=\frac{1}{n}\sum_{t=1}^{\smlfloor{n\lambda}}z_{n,t-m}^{d}\indic^{+}(y_{t-m})-\frac{1}{n}\sum_{t=1}^{\smlfloor{n\lambda}}z_{n,t-m}^{d}\indic\{0\leq y_{t-m}<n^{1/2}\delta\}
\]
and by \lemref{YZwkc}\ref{enu:YZwkc:ydelta},
\begin{align*}
\norm{\frac{1}{n}\sum_{t=1}^{\smlfloor{n\lambda}}z_{n,t-m}^{d}\indic\{0\leq y_{t-m}<n^{1/2}\delta\}} & \leq\max_{s\leq n}\smlnorm{z_{n,s}^{d}}\frac{1}{n}\sum_{t=1}^{n}\indic\{\smlabs{y_{t-m}}<n^{1/2}\delta\}\\
 & \leq\max_{s\leq n}\smlnorm{z_{n,s}^{d}}\left(\frac{1}{n}\sum_{t=1}^{n}\indic\{\smlabs{y_{t}}<n^{1/2}\delta\}+o_{p}(1)\right)\inprob0
\end{align*}
as $n\goesto\infty$, $m\goesto\infty$ and then $\delta\goesto0$.
Hence by \lemref{YZwkc}\ref{enu:YZwkc:YZ},
\[
\frac{1}{n}\sum_{t=1}^{\smlfloor{n\lambda}}z_{n,t-m}^{d}\indic\{y_{t-m}\geq n^{1/2}\delta\}=\frac{1}{n}\sum_{t=1}^{\smlfloor{n\lambda}}z_{n,t}^{d}\indic^{+}(y_{t})+o_{p}(1)\wkc\int_{0}^{\lambda}\indic^{+}[Y(s)]Z(s)\diff s.
\]
as $n\goesto\infty$, $m\goesto\infty$ and then $\delta\goesto0$.
Since $g$ is bounded and continuous, and
\[
w_{m,0}^{+}=\sum_{\ell=0}^{m-1}(A^{+})^{\ell}(c^{+}+B^{+}v_{-\ell})\inas\sum_{\ell=0}^{\infty}(A^{+})^{\ell}(c^{+}+B^{+}v_{-\ell})=w_{0}^{+},
\]
it follows by dominated convergence theorem that $\expect g(w_{m,0}^{+})\goesto\expect g(w_{0}^{+})$
as $m\goesto\infty$. Hence
\[
\expect g(w_{m,0}^{+})\frac{1}{n}\sum_{t=1}^{\smlfloor{n\lambda}}z_{n,t-m}^{d}\indic\{y_{t-m}\geq n^{1/2}\delta\}\wkc\expect g(w_{0}^{+})\int_{0}^{\lambda}\indic^{+}[Y(s)]Z(s)\diff s
\]
as $n\goesto\infty$, $m\goesto\infty$ and then $\delta\goesto0$.
\end{proof}
\begin{proof}[Proof of \lemref{comp}]
 \textbf{\ref{enu:comp:nonsingular}.} Recall from \eqref{beta-ast}
and \eqref{tau-ast} that
\begin{align*}
\ctr^{\ast} & =\begin{bmatrix}1 & 0 & \ctr_{xy}^{+\trans}\\
0 & 1 & \ctr_{xy}^{-\trans}\\
0 & 0 & \beta_{x,\perp}
\end{bmatrix} & \beta^{\ast} & =\begin{bmatrix}\beta_{y}^{+\trans}\\
\beta_{y}^{-\trans}\\
\beta_{x}
\end{bmatrix}.
\end{align*}
Let $a=(a_{1},a_{2},a_{(3)}^{\trans})^{\trans}\in\reals^{q+1}$ and
$b\in\reals^{r}$ be such that
\[
0=\begin{bmatrix}\tau^{\ast} & \beta^{\ast}\end{bmatrix}\begin{bmatrix}a\\
b
\end{bmatrix}=\begin{bmatrix}a_{1}+\ctr_{xy}^{+\trans}a_{(3)}+\beta_{y}^{+\trans}b\\
a_{2}+\ctr_{xy}^{-\trans}a_{(3)}+\beta_{y}^{-\trans}b\\
\beta_{x,\perp}a_{(3)}+\beta_{x}b
\end{bmatrix}
\]
where $a_{(3)}\in\reals^{q-1}$. Since $[\beta_{x,\perp},\beta_{x}]$
has rank $p-1$, it follows that $a_{(3)}=0$ and $b=0$. Hence $a_{1}=a_{2}=0$,
i.e.\ $a=0$ also.

\textbf{\ref{enu:comp:limits}.} Regarding $\dmn{\varrho}_{n,t}$,
we have by \eqref{rhocvg} that
\begin{align*}
\frac{1}{n}\sum_{t=1}^{\smlfloor{n\lambda}}\dmn{\varrho}_{n,t} & =\frac{1}{n}\sum_{t=1}^{\smlfloor{n\lambda}}\varrho_{n,t}-\lambda\frac{1}{n}\sum_{t=1}^{n}\varrho_{n,t}\\
 & \wkc\int_{0}^{\lambda}R(s)\diff s-\lambda\int_{0}^{1}R(s)\diff s=\int_{0}^{\lambda}\dmn R(s)\diff s
\end{align*}
on $D[0,1]$ jointly with $U_{n}\wkc U$. We next consider $\dmn{\xi}_{t}$,
for which we similarly have
\begin{align}
\frac{1}{n}\sum_{t=1}^{\smlfloor{n\lambda}}\dmn{\xi}_{t} & =\frac{1}{n}\sum_{t=1}^{\smlfloor{n\lambda}}\xi_{t}-\lambda\frac{1}{n}\sum_{t=1}^{n}\xi_{t}.\label{eq:xitildeavg}
\end{align}
To determine the weak limits of the various components on the r.h.s.,\ we
apply \lemref{pmavg}. To that end, define 
\[
\b{\xi}_{t}\defeq\b{\beta}(y_{t})^{\trans}\b z_{t}=(\xi_{t}^{\trans},\Delta z_{t}^{\ast\trans},\ldots,\Delta z_{t-k+2}^{\ast\trans})^{\trans}
\]
where as per \eqref{alpabeta},
\begin{align}
\b{\alpha}\defeq & \begin{bmatrix}\alpha & \Gamma_{1} & \Gamma_{2} & \cdots & \Gamma_{k-1}\\
 & I_{p+1}\\
 &  & I_{p+1}\\
 &  &  & \ddots\\
 &  &  &  & I_{p+1}
\end{bmatrix}, & \b{\beta}(y)^{\trans} & \defeq\begin{bmatrix}\beta(y)^{\trans}\\
S_{p}(y) & -I_{p+1}\\
 & I_{p+1} & -I_{p+1}\\
 &  & \ddots & \ddots\\
 &  &  & I_{p+1} & -I_{p+1}
\end{bmatrix},\label{eq:balphabbeta}
\end{align}
and
\begin{align}
\b c & \defeq\begin{bmatrix}c\\
0_{(p+1)(k-1)}
\end{bmatrix} & \b u_{t} & \defeq\begin{bmatrix}u_{t}\\
0_{(p+1)(k-1)}
\end{bmatrix}\label{eq:bcu}
\end{align}
it follows by Lemma~B.2 and the arguments subsequently given in the
proof of Theorem~4.2 in \DMW{}, that $w_{t}=\b{\xi}_{t}$ follows
an autoregressive process satisfying the requirements of \lemref{pmavg}
above (see the statement of \thmref{lln}), with in particular
\begin{align*}
c^{\pm} & =\b{\beta}(\pm1)^{\trans}\b c, & A^{\pm} & =I_{r+(k-1)(p+1)}+\b{\beta}(\pm1)^{\trans}\b{\alpha}, & B^{\pm} & =\b{\beta}(\pm1)^{\trans}, & v_{t} & =\b u_{t}.
\end{align*}
Hence by that result, with $g(w)=w$ and noting that $\smlnorm{v_{t}}_{2+\delta_{u}}<\infty$,
\begin{align}
\frac{1}{n}\sum_{t=1}^{\smlfloor{n\lambda}}\xi_{t}=E_{r}^{\trans}\frac{1}{n}\sum_{t=1}^{\smlfloor{n\lambda}}\b{\xi}_{t} & =E_{r}^{\trans}\left[\frac{1}{n}\sum_{t=1}^{\smlfloor{n\lambda}}\b{\xi}_{t}\indic^{+}(y_{t})+\frac{1}{n}\sum_{t=1}^{\smlfloor{n\lambda}}\b{\xi}_{t}\indic^{-}(y_{t})\right]\nonumber \\
 & \wkc E_{r}^{\trans}[(\expect\b{\xi}_{0}^{+})m_{Y}^{+}(\lambda)+(\expect\b{\xi}_{0}^{-})m_{Y}^{-}(\lambda)]\nonumber \\
 & =\mu_{\xi}^{+}m_{Y}^{+}(\lambda)+\mu_{\xi}^{-}m_{Y}^{-}(\lambda)=\lambda\mu_{\xi},\label{eq:xiavg}
\end{align}
where $E_{r}$ denotes the first $r$ columns of $I_{r+(k-1)(p+1)}$,
\begin{equation}
\b{\xi}_{0}^{\pm}\defeq-[\b{\beta}(\pm1)^{\trans}\b{\alpha}]^{-1}\b{\beta}(\pm1)^{\trans}\b c+\sum_{\ell=0}^{\infty}[I_{r+(k-1)(p+1)}+\b{\beta}(\pm1)^{\trans}\b{\alpha}]^{\ell}\b{\beta}(\pm1)^{\trans}\b u_{-\ell},\label{eq:bxi0-def}
\end{equation}
and for $\xi_{0}^{\pm}\defeq E_{r}^{\trans}\b{\xi}_{0}^{\pm}$,
\begin{equation}
\mu_{\xi}^{\pm}\defeq\expect\xi_{0}^{\pm}=-E_{r}^{\trans}[\b{\beta}(\pm1)^{\trans}\b{\alpha}]^{-1}\b{\beta}(\pm1)^{\trans}\b c=\mu_{\xi}\label{eq:muxi-def}
\end{equation}
because by \assref{det} there exists a $\mu_{\xi}\in\reals^{r}$
such that $c=-\alpha\mu_{\xi}$, and therefore $\b c=-\b{\alpha}\b{\mu}_{\xi}$
for $\b{\mu}_{\xi}\defeq(\mu_{\xi}^{\trans},0_{(p+1)(k-1)}^{\trans})^{\trans}$.
Hence it follows from \eqref{xitildeavg} and \eqref{xiavg} that
\begin{equation}
\frac{1}{n}\sum_{t=1}^{\smlfloor{n\lambda}}\dmn{\xi}_{t}=\frac{1}{n}\sum_{t=1}^{\smlfloor{n\lambda}}\xi_{t}-\lambda\frac{1}{n}\sum_{t=1}^{n}\xi_{t}\inprob\lambda\mu_{\xi}-\lambda\mu_{\xi}=0.\label{eq:xicvg}
\end{equation}
on $D[0,1]$.

\textbf{\ref{enu:comp:second}.} Observe that because $\dmn{\varrho}_{n,t}$
and $\dmn{\xi}_{t}$ have zero sample mean,
\begin{equation}
\frac{1}{n}\sum_{t=1}^{n}\begin{bmatrix}\dmn{\varrho}_{n,t}\dmn{\varrho}_{n,t}^{\trans} & \dmn{\varrho}_{n,t}\dmn{\xi}_{t}^{\trans}\\
\dmn{\xi}_{t}\dmn{\varrho}_{n,t}^{\trans} & \dmn{\xi}_{t}\dmn{\xi}_{t}^{\trans}
\end{bmatrix}=\frac{1}{n}\sum_{t=1}^{n}\begin{bmatrix}\dmn{\varrho}_{n,t}\varrho_{n,t}^{\trans} & \varrho_{n,t}\dmn{\xi}_{t}^{\trans}\\
\dmn{\xi}_{t}\varrho_{n,t}^{\trans} & \dmn{\xi}_{t}\xi_{t}^{\trans}
\end{bmatrix}.\label{eq:afterorthog}
\end{equation}
For the upper left block of \eqref{afterorthog}, we have directly
from \eqref{rhocvg} that
\begin{align*}
\frac{1}{n}\sum_{t=1}^{n}\dmn{\varrho}_{n,t}\varrho_{n,t}^{\trans} & =\frac{1}{n}\sum_{t=1}^{n}\varrho_{n,t}\varrho_{n,t}^{\trans}-\hat{\mu}_{n,\varrho}\hat{\mu}_{n,\varrho}^{\trans}\\
 & \wkc\int_{0}^{1}R(s)R(s)^{\trans}\diff s-\left(\int_{0}^{1}R(s)\diff s\right)\left(\int_{0}^{1}R(s)\diff s\right)^{\trans}\\
 & =\int_{0}^{1}\dmn R(s)\dmn R(s)^{\trans}\diff s,
\end{align*}
where $\hat{\mu}_{n,\varrho}=\frac{1}{n}\sum_{t=1}^{\smlfloor{n\lambda}}\varrho_{n,t}$.

We next consider the off-diagonal block, for which 
\begin{align*}
\frac{1}{n}\sum_{t=1}^{n}\dmn{\xi}_{t}\varrho_{n,t}^{\trans} & =\frac{1}{n}\sum_{t=1}^{n}(\xi_{t}-\hat{\mu}_{n,\xi})\varrho_{n,t}^{\trans}\\
 & =\frac{1}{n}\sum_{t=1}^{n}(\xi_{t}-\hat{\mu}_{n,\xi})\varrho_{n,t}^{\trans}\indic^{+}(y_{t})+\frac{1}{n}\sum_{t=1}^{n}(\xi_{t}-\hat{\mu}_{n,\xi})\varrho_{n,t}^{\trans}\indic^{-}(y_{t})
\end{align*}
since $\indic^{+}(y_{t})+\indic^{-}(y_{t})=1$, where $\hat{\mu}_{n,\xi}=\frac{1}{n}\sum_{t=1}^{n}\xi_{t}$.
Using, as noted in the proof of part~\ref{enu:comp:limits}, that
$\xi_{t}=E_{r}^{\trans}\b{\xi}_{t}$, it follows from \eqref{rhocvg}
and \eqref{Zastavg} (itself an implication of \lemref{pmavg}) and
\eqref{muxi-def} that
\begin{align*}
\frac{1}{n}\sum_{t=1}^{n}\xi_{t}\varrho_{n,t}^{\trans}\indic^{\pm}(y_{t}) & =E_{r}^{\trans}\left[\frac{1}{n^{3/2}}\sum_{t=1}^{n}\indic^{\pm}(y_{t})\b{\xi}_{t}z_{t}^{\ast\trans}\right]\ctr^{\ast}\\
 & \wkc E_{r}^{\trans}[\expect\b{\xi}_{0}^{\pm}]\left[\int_{0}^{1}Z^{\ast}(s)\indic^{\pm}[Y(s)]\diff s\right]^{\trans}\ctr^{\ast}\\
 & =(\expect\xi_{0}^{\pm})\int_{0}^{1}R(s)^{\trans}\indic^{\pm}[Y(s)]\diff s\\
 & =\mu_{\xi}\int_{0}^{1}R(s)^{\trans}\indic^{\pm}[Y(s)]\diff s
\end{align*}
while by another application of \lemref{pmavg}, and \eqref{xiavg}
above (with $\lambda=1$)
\[
\hat{\mu}_{n,\xi}\frac{1}{n}\sum_{t=1}^{n}\varrho_{n,t}^{\trans}\indic^{\pm}(y_{t})\wkc\mu_{\xi}\int_{0}^{1}R(s)^{\trans}\indic^{\pm}[Y(s)]\diff s.
\]
Deduce that
\[
\frac{1}{n}\sum_{t=1}^{n}(\xi_{t}-\hat{\mu}_{n,\xi})\varrho_{n,t}^{\trans}\indic^{\pm}(y_{t})\inprob0,
\]
and thus $\frac{1}{n}\sum_{t=1}^{n}\dmn{\xi}_{t}\varrho_{n,t}^{\trans}\inprob0$,
as required.

We come finally to the lower right block of \eqref{afterorthog}.
We have
\begin{align}
\frac{1}{n}\sum_{t=1}^{n}\dmn{\xi}_{t}\xi_{t}^{\trans} & =\frac{1}{n}\sum_{t=1}^{n}(\xi_{t}-\hat{\mu}_{n,\xi})\xi_{t}^{\trans}=\frac{1}{n}\sum_{t=1}^{n}\xi_{t}\xi_{t}^{\trans}-\hat{\mu}_{n,\xi}\hat{\mu}_{n,\xi}^{\trans}\label{eq:xixi}
\end{align}
where $\hat{\mu}_{n,\xi}\inprob\mu_{\xi}$ per \eqref{xiavg} above.
Similarly to \eqref{xiavg}, we also have by \lemref{pmavg} (in this
instance with $g(w)=ww^{\trans}$, and noting that $\smlnorm{v_{t}}_{2+\delta_{u}}<\infty$)
that
\begin{equation}
\frac{1}{n}\sum_{t=1}^{n}\xi_{t}\xi_{t}^{\trans}\indic^{\pm}(y_{t})=E_{r}^{\trans}\left[\frac{1}{n}\sum_{t=1}^{n}\b{\xi}_{t}\b{\xi}_{t}^{\trans}\indic^{\pm}(y_{t})\right]E_{r}\wkc(\expect\xi_{0}^{\pm}\xi_{0}^{\pm\trans})m_{Y}^{\pm}(1).\label{eq:xixipm}
\end{equation}
By \eqref{bxi0-def} and \eqref{muxi-def} above,
\begin{align*}
\xi_{0}^{\pm}-\mu_{\xi}=E_{r}^{\trans}[\b{\xi}_{0}^{\pm}-\expect\b{\xi}_{0}^{\pm}] & =E_{r}^{\trans}\sum_{\ell=0}^{\infty}(I_{r+(k-1)(p+1)}+\b{\beta}(\pm1)^{\trans}\b{\alpha})^{\ell}\b{\beta}(\pm1)^{\trans}\b u_{-\ell}.
\end{align*}
Recalling the definitions of $\b{\beta}(y)$ and $\b u_{t}$ in \eqref{balphabbeta}
and \eqref{bcu} above, the first term on the r.h.s.\ series is
\[
E_{r}^{\trans}\b{\beta}(\pm1)^{\trans}\b u_{0}=\beta(\pm1)^{\trans}u_{0},
\]
which has nonsingular matrix variance $\beta(\pm1)^{\trans}\Sigma_{u}\beta(\pm1)$.
It follows that $\Sigma_{\xi}^{\pm}\defeq\var(\xi_{0}^{\pm})$ is
positive definite, and since
\[
\expect\xi_{0}^{\pm}\xi_{0}^{\pm\trans}=\Sigma_{\xi}^{\pm}+\mu_{\xi}\mu_{\xi}^{\trans}
\]
we deduce from \eqref{xixi} and \eqref{xixipm} that
\begin{align*}
\frac{1}{n}\sum_{t=1}^{n}\dmn{\xi}_{t}\xi_{t}^{\trans} & \wkc(\Sigma_{\xi}^{+}+\mu_{\xi}\mu_{\xi}^{\trans})m_{Y}^{+}(1)+(\Sigma_{\xi}^{-}+\mu_{\xi}\mu_{\xi}^{\trans})m_{Y}^{-}(1)-\mu_{\xi}\mu_{\xi}^{\trans}\\
 & =\Sigma_{\xi}^{+}m_{Y}^{+}(1)+\Sigma_{\xi}^{-}m_{Y}^{-}(1).
\end{align*}
Since $m_{Y}^{+}(1)+m_{Y}^{-}(1)=1$, this is positive definite as
the convex combination of two positive definite matrices.
\end{proof}
\begin{proof}[Proof of \lemref{stdised}]
 In view of \eqref{zd-cvg}, \eqref{zastcvg} and \eqref{rhocvg},
we have
\begin{equation}
R(\lambda)=\ctr^{\ast\trans}Z^{\ast}(\lambda)=\ctr^{\ast\trans}S_{p}[Y(\lambda)]Z(\lambda)=\ctr^{\ast\trans}S_{p}[Y(\lambda)]P_{\beta_{\perp}}[Y(\lambda)]U_{0}(\lambda).\label{eq:Rdef}
\end{equation}
As in Lemma~B.3 in \DMW{}, define $g(y,u)\defeq P_{\beta_{\perp}}(y)u$
and $\vartheta^{\trans}\defeq e_{1}^{\trans}P_{\beta_{\perp}}(+1)\neq0$.
It follows from Theorem~4.2 in \DMW{} that $\sgn Y(\lambda)=\sgn\vartheta^{\trans}U_{0}(\lambda)$,
and therefore
\begin{equation}
Z^{\ast}(\lambda)=S_{p}[Y(\lambda)]P_{\beta_{\perp}}[Y(\lambda)]U_{0}(\lambda)=S_{p}[\vartheta^{\trans}U_{0}(\lambda)]P_{\beta_{\perp}}[\vartheta^{\trans}U_{0}(\lambda)]U_{0}(\lambda).\label{eq:Zexp}
\end{equation}
The r.h.s.\ is a (continuous) function of a $p$-dimensional Brownian
motion $U_{0}(\lambda)$; our objective is to rewrite it in terms
of a (known) function of a $q$-dimensional standard (up to initialisation)
Brownian motion $W_{0}$. The chief obstacle here (relative to the
linear case) lies in the nonlinearity with which $U_{0}$ enters the
r.h.s.; we therefore first seek to obtain a expression for $Z^{\ast}$
in terms of a $p$-dimensional Brownian motion, such that only the
first component of that Brownian motion enters $Z^{\ast}(\lambda)$
nonlinearly.

To that end, define $\theta\defeq\smlnorm{\vartheta}^{-1}\vartheta$,
and let $\Theta\defeq[\theta,\theta_{\perp}]$ be a $p\times p$ orthonormal
matrix. Then for any $y\in\reals$ and $u\in\reals^{p}$,
\begin{align*}
g(y,u) & =P_{\beta_{\perp}}(y)u=P_{\beta_{\perp}}(y)\Theta\Theta^{\trans}u=\begin{bmatrix}P_{\beta_{\perp}}(y)\theta & P_{\beta_{\perp}}(y)\theta_{\perp}\end{bmatrix}\begin{bmatrix}\theta^{\trans}u\\
\theta_{\perp}^{\trans}u
\end{bmatrix},
\end{align*}
and note that $\vartheta^{\trans}\theta_{\perp}=0$ by construction.
Therefore applying Lemma~B.3(ii) in \DMW{} to each column of $P_{\beta_{\perp}}(y)\theta_{\perp}$,
we obtain 
\[
P_{\beta_{\perp}}(+1)\theta_{\perp}=P_{\beta_{\perp}}(-1)\theta_{\perp}
\]
whence
\[
g(y,u)=P_{\beta_{\perp}}(y)\theta[\theta^{\trans}u]+P_{\beta_{\perp}}(+1)\theta_{\perp}[\theta_{\perp}^{\trans}u].
\]
This allows us to confine the nonlinearity in the function to the
scalar variable $\theta^{\trans}u$, with the remaining $p-1$ variables
$\theta_{\perp}^{\trans}u$ entering the r.h.s.\ linearly. In view
of \eqref{Zexp}, which because $\sgn\vartheta^{\trans}u=\sgn\theta^{\trans}u$
may be written as
\begin{equation}
Z^{\ast}(\lambda)=S_{p}[\theta^{\trans}U_{0}(\lambda)]P_{\beta_{\perp}}[\theta^{\trans}U_{0}(\lambda)]U_{0}(\lambda)=S_{p}[\theta^{\trans}U_{0}(\lambda)]g[\theta^{\trans}U_{0}(\lambda),U_{0}(\lambda)],\label{eq:Zrep}
\end{equation}
we are only interested in the case where $\sgn y=\sgn\theta^{\trans}u$,
for which
\begin{align}
g(\theta^{\trans}u,u) & =P_{\beta_{\perp}}(\theta^{\trans}u)\theta[\theta^{\trans}u]+P_{\beta_{\perp}}(+1)\theta_{\perp}[\theta_{\perp}^{\trans}u]\nonumber \\
 & =P_{\beta_{\perp}}(+1)\theta[\theta^{\trans}u]_{+}+P_{\beta_{\perp}}(-1)\theta[\theta^{\trans}u]_{-}+P_{\beta_{\perp}}(+1)\theta_{\perp}[\theta_{\perp}^{\trans}u]\nonumber \\
 & \eqdef\psi^{+}[\theta^{\trans}u]_{+}+\psi^{-}[\theta^{\trans}u]_{-}+\Psi^{x}[\theta_{\perp}^{\trans}u].\label{eq:grep}
\end{align}
By Lemma~B.3(i) in \DMW{},
\[
e_{1}^{\trans}\psi^{+}=e_{1}^{\trans}P_{\beta_{\perp}}(+1)\theta=\frac{\vartheta^{\trans}\vartheta}{\smlnorm{\vartheta}}=\smlnorm{\vartheta}>0
\]
and also $e_{1}^{\trans}\psi^{-}>0$, while
\begin{equation}
e_{1}^{\trans}\Psi^{x}=e_{1}^{\trans}P_{\beta_{\perp}}(+1)\theta_{\perp}=\vartheta^{\trans}\theta_{\perp}=0,\label{eq:psixorthog}
\end{equation}
and thus we may write $\Psi^{x}=[0_{q-1}^{\trans},\Psi_{xx}^{\trans}]^{\trans}$
for some $\Psi_{xx}\in\reals^{(p-1)\times(q-1)}$. Partitioning $\psi^{\pm}=(\psi_{y}^{\pm},\psi_{x}^{\pm\trans})^{\trans}$,
where $\psi_{y}^{\pm}\defeq e_{1}^{\trans}\psi^{\pm}$, we obtain
from \eqref{Zrep} and \eqref{grep} the representation
\begin{align}
Z^{\ast}(\lambda) & =\begin{bmatrix}\indic^{+}[\theta^{\trans}U_{0}(\lambda)] & 0\\
\indic^{-}[\theta^{\trans}U_{0}(\lambda)] & 0\\
0 & I_{p-1}
\end{bmatrix}\begin{bmatrix}\psi_{y}^{+} & \psi_{y}^{-} & 0\\
\psi_{x}^{+} & \psi_{x}^{-} & \Psi_{xx}
\end{bmatrix}\begin{bmatrix}[\theta^{\trans}U_{0}(\lambda)]_{+}\\{}
[\theta^{\trans}U_{0}(\lambda)]_{-}\\
\theta_{\perp}^{\trans}U_{0}(\lambda)
\end{bmatrix}\nonumber \\
 & =\begin{bmatrix}\psi_{y}^{+} & 0 & 0\\
0 & \psi_{y}^{-} & 0\\
\psi_{x}^{+} & \psi_{x}^{-} & \Psi_{xx}
\end{bmatrix}\begin{bmatrix}[\theta^{\trans}U_{0}(\lambda)]_{+}\\{}
[\theta^{\trans}U_{0}(\lambda)]_{-}\\
\theta_{\perp}^{\trans}U_{0}(\lambda)
\end{bmatrix}\eqdef\Psi^{\ast}S_{p}[\theta^{\trans}U_{0}(\lambda)]\Theta^{\trans}U_{0}(\lambda)\label{eq:Z2rep}
\end{align}
where we have used the fact that $\psi_{y}^{\pm}>0$. We have thus
represented $Z^{\ast}$ in terms of a $p$-dimensional Brownian motion
$\Theta^{\trans}U_{0}$, where only the first component $e_{1}^{\trans}\Theta^{\trans}U_{0}(\lambda)=\theta^{\trans}U_{0}(\lambda)$
enters $Z^{\ast}(\lambda)$ nonlinearly.

The next step is to collapse the $(p+1)$-dimensional process $Z^{\ast}(\lambda)$
into the $(q+1)$-dimensional process $R(\lambda)$. From \eqref{Rdef}
and \eqref{Z2rep}, we have
\[
R(\lambda)=\ctr^{\ast\trans}Z^{\ast}(\lambda)=\begin{bmatrix}1 & 0 & 0\\
0 & 1 & 0\\
\ctr_{xy}^{+} & \ctr_{xy}^{-} & \beta_{x,\perp}^{\trans}
\end{bmatrix}\begin{bmatrix}\psi_{y}^{+} & 0 & 0\\
0 & \psi_{y}^{-} & 0\\
\psi_{x}^{+} & \psi_{x}^{-} & \Psi_{xx}
\end{bmatrix}S_{p}[\theta^{\trans}U_{0}(\lambda)]\Theta^{\trans}U_{0}(\lambda)
\]
where we are entirely free to choose $\ctr_{xy}^{\pm}\in\reals^{q-1}$,
in view of \lemref{comp}. (Note that the corresponding choice of
$\ctr_{xy}^{\pm}$ is then embedded into the definition of $R(\lambda)$.)
In particular, if we take
\[
\ctr_{xy}^{\pm}\defeq-\beta_{x,\perp}^{\trans}\psi_{x}^{\pm}(\psi_{y}^{\pm})^{-1},
\]
as is permitted since $\psi_{y}^{\pm}\neq0$, then it will follow
that 
\begin{align*}
R(\lambda)=\ctr^{\ast\trans}\Psi^{\ast}\begin{bmatrix}[\theta^{\trans}U_{0}(\lambda)]_{+}\\{}
[\theta^{\trans}U_{0}(\lambda)]_{-}\\
\theta_{\perp}^{\trans}U_{0}(\lambda)
\end{bmatrix} & =\begin{bmatrix}\psi_{y}^{+} & 0 & 0\\
0 & \psi_{y}^{-} & 0\\
0 & 0 & \beta_{x,\perp}^{\trans}\Psi_{xx}
\end{bmatrix}\begin{bmatrix}[\theta^{\trans}U_{0}(\lambda)]_{+}\\{}
[\theta^{\trans}U_{0}(\lambda)]_{-}\\
\theta_{\perp}^{\trans}U_{0}(\lambda)
\end{bmatrix}\\
 & =\begin{bmatrix}\psi_{y}^{+} & 0 & 0\\
0 & \psi_{y}^{-} & 0\\
0 & 0 & I_{q-1}
\end{bmatrix}\begin{bmatrix}[\theta^{\trans}U_{0}(\lambda)]_{+}\\{}
[\theta^{\trans}U_{0}(\lambda)]_{-}\\
\beta_{x,\perp}^{\trans}\Psi_{xx}\theta_{\perp}^{\trans}U_{0}(\lambda)
\end{bmatrix}.
\end{align*}
Defining
\[
B_{0}(\lambda)\defeq\begin{bmatrix}\theta^{\trans}\\
\beta_{x,\perp}^{\trans}\Psi_{xx}\theta_{\perp}^{\trans}
\end{bmatrix}U_{0}(\lambda)
\]
we thus obtain a $q$-dimensional Brownian motion. To show that it
has full rank variance matrix, since $\theta\neq0$ and $\beta_{x,\perp}^{\trans}\Psi_{xx}\theta_{\perp}^{\trans}\theta=0$,
it suffices to show that $\rank\beta_{x,\perp}^{\trans}\Psi_{xx}=q-1$.

To that end, we first note that by the remark following \eqref{psixorthog}
above,
\begin{equation}
\beta_{x,\perp}^{\trans}\Psi_{xx}=[0_{q-1},\beta_{x,\perp}^{\trans}]\Psi^{x}=[0_{q-1},\beta_{x,\perp}^{\trans}]P_{\beta_{\perp}}(+1)\theta_{\perp},\label{eq:betapsi}
\end{equation}
The columns of $\Psi^{x}=P_{\beta_{\perp}}(+1)\theta_{\perp}$ are
orthogonal to those of $e_{p,1}$ (by \eqref{psixorthog} above) and
of $\beta(+1)$; while $\rank[e_{p,1},\beta(+1)]=r+1$, because $e_{p,1}^{\trans}\beta_{\perp}(+1)\neq0$
(by Lemma~B.3(i) in \DMW{}) and so cannot be contained in the span
of $\beta(+1)$. It follows that the ($p-1$) columns of $\Psi^{x}$
span the $(q-1)$-dimensional subspace of $\reals^{p}$ that is orthogonal
to $[e_{p,1},\beta(+1)]$. Since the ($q-1$) columns of 
\[
\begin{bmatrix}0_{q-1}^{\trans}\\
\beta_{x,\perp}
\end{bmatrix}\in\reals^{p\times(q-1)}
\]
also span that subspace, it follows from \eqref{betapsi} that $[0_{q-1},\beta_{x,\perp}^{\trans}]\Psi^{x}=\beta_{x,\perp}^{\trans}\Psi_{xx}$
has rank $q-1$. Letting $D_{\psi}\defeq\diag\{\psi_{y}^{+},\psi_{y}^{-},I_{q-1}\}$,
we have thus obtained
\[
R(\lambda)=D_{\psi}S_{q}[e_{1}^{\trans}B_{0}(\lambda)]B_{0}(\lambda),
\]
where $B_{0}$ is a $q$-dimensional Brownian motion.

The final step is to recognise that, despite the nonlinearity on the
r.h.s., we may still render this in terms of a standard (up to initialisation)
Brownian motion by means of the usual Cholesky factorisation. Let
$\Sigma_{B}$ denote the variance of $B_{0}$, and let $L$ denote
the (lower triangular) Cholesky root of $\Sigma_{B}^{-1}$, so that
\[
W_{0}(\lambda)\defeq LB_{0}(\lambda)
\]
is a $q$-dimensional standard (up to initialisation) Brownian motion.
Partitioning $L$ and defining $L^{\ast}$ as
\begin{align*}
L & =\begin{bmatrix}\ell_{1} & 0\\
\ell_{(2),1} & L_{(2)}
\end{bmatrix}, & L^{\ast} & \defeq\begin{bmatrix}\ell_{1} & 0 & 0\\
0 & \ell_{1} & 0\\
\ell_{(2),1} & \ell_{(2),1} & L_{(2)}
\end{bmatrix},
\end{align*}
where $\ell_{(1)}>0$ is scalar, and $L_{(2)}\in\reals^{(q-1)\times(q-1)}$,
and partitioning $I_{q}=[e_{q,1},E_{q,-1}]$, we obtain
\begin{align}
L^{\ast}D_{\psi}^{-1}R(\lambda) & =L^{\ast}S_{q}[e_{1}^{\trans}B_{0}(\lambda)]B_{0}(\lambda)\nonumber \\
 & =\begin{bmatrix}\ell_{1} & 0 & 0\\
0 & \ell_{1} & 0\\
\ell_{(2),1} & \ell_{(2),1} & L_{(2)}
\end{bmatrix}\begin{bmatrix}[e_{q,1}^{\trans}B_{0}(\lambda)]_{+}\\{}
[e_{q,1}^{\trans}B_{0}(\lambda)]_{-}\\
E_{q,-1}^{\trans}B_{0}(\lambda)
\end{bmatrix}=\begin{bmatrix}[\ell_{1}e_{q,1}^{\trans}B_{0}(\lambda)]_{+}\\{}
[\ell_{1}e_{q,1}^{\trans}B_{0}(\lambda)]_{-}\\
(\ell_{(2),1}e_{q,1}^{\trans}+L_{(2)}E_{q,-1}^{\trans})B_{0}(\lambda)
\end{bmatrix}\nonumber \\
 & =\begin{bmatrix}[e_{q,1}^{\trans}W_{0}(\lambda)]_{+}\\{}
[e_{q,1}^{\trans}W_{0}(\lambda)]_{-}\\
E_{q,-1}^{\trans}W_{0}(\lambda)
\end{bmatrix}=S_{q}[e_{q,1}^{\trans}W_{0}(\lambda)]W_{0}(\lambda)=W_{0}^{\ast}(\lambda).\label{eq:Rrep}
\end{align}
Hence the result for $R$ holds with $Q=L^{\ast}D_{\psi}^{-1}$.

To obtain the desired representation for $Y$, we first invert \eqref{Rrep}
to write
\[
\ctr^{\ast\trans}Z^{\ast}(\lambda)=R(\lambda)=D_{\psi}(L^{\ast})^{-1}W_{0}^{\ast}(\lambda).
\]
Let $E_{d,2}$ denote the first two columns of $I_{d}$. Because the
first two rows of each of $(L^{\ast})^{-1}$, $D_{\psi}$ and $\ctr^{\ast\trans}$
are zero everywhere except for the $(1,1)$ and $(2,2)$ elements,
we have
\[
E_{q+1,2}^{\trans}D_{\psi}(L^{\ast})^{-1}=\begin{bmatrix}\ell_{1}^{-1}\psi_{y}^{+} & 0 & 0_{1\times(q-1)}\\
0 & \ell_{1}^{-1}\psi_{y}^{-} & 0_{1\times(q-1)}
\end{bmatrix}
\]
and $E_{q+1,2}^{\trans}\ctr^{\ast\trans}=E_{p+1,2}^{\trans}$. Hence
\[
\begin{bmatrix}[Y(\lambda)]_{+}\\{}
[Y(\lambda)]_{-}
\end{bmatrix}=E_{p+1,2}^{\trans}Z^{\ast}(\lambda)=E_{q+1,2}^{\trans}\ctr^{\ast\trans}Z^{\ast}(\lambda)=E_{q+1,2}^{\trans}R(\lambda)=\begin{bmatrix}\ell_{1}^{-1}\psi_{y}^{+}[e_{q,1}^{\trans}W_{0}(\lambda)]_{+}\\
\ell_{1}^{-1}\psi_{y}^{-}[e_{q,1}^{\trans}W_{0}(\lambda)]_{-}
\end{bmatrix}
\]
whence the claim follows with $\omega^{\pm}=\ell_{1}^{-1}\psi_{y}^{\pm}>0$.
\end{proof}
\begin{proof}[Proof of \lemref{rank}]
 Since $\mathcal{W}_{0}=0$, we have $W_{0}=W$, a $q$-dimensional
standard Brownian motion (initialised at zero). To reduce the notational
clutter, we will drop the `$0$' subscript from $\dmn W_{0}^{\ast}$
and $\dmn V_{0}^{\ast}$ throughout what follows.

We first consider $\dmn S_{V}^{\ast}$. We note that a realisation
of the positive semi-definite matrix $\dmn S_{V}^{\ast}$ is rank
deficient if and only if there exists (for that realisation) an $a\in\reals^{q+1}$
such that
\[
0=a^{\trans}\dmn S_{V}^{\ast}a=\int_{0}^{1}[a^{\trans}\dmn V^{\ast}(s)]^{2}\diff s.
\]
Since $\dmn V^{\ast}(s)=\int_{0}^{s}\dmn W^{\ast}(\lambda)\diff\lambda$
has continuous paths, the preceding implies that
\[
0=a^{\trans}\dmn V^{\ast}(s)=a^{\trans}\int_{0}^{s}\dmn W^{\ast}(\lambda)\diff\lambda
\]
for all $s\in[0,1]$; and hence, differentiating with respect to $s$,
that
\[
a^{\trans}\dmn W^{\ast}(\lambda)=0
\]
for all $\lambda\in[0,1]$. Since $\dmn W^{\ast}$ itself has continuous
paths, a realisation of $\dmn S_{W}^{\ast}$ is rank deficient only
if there exists an $a$ such that the preceding condition holds. Hence
it suffices to show that
\begin{equation}
\Prob\{\exists a\in\reals^{q+1}\text{ s.t. }a^{\trans}\dmn W^{\ast}(\lambda)=0\sep\forall\lambda\in[0,1]\}=0.\label{eq:Wrkdef}
\end{equation}

Since $\dmn W^{\ast}$ is the residual from an $L^{2}([0,1])$ projection
of (each element of) the $(q+1)$-dimensional process
\[
W^{\ast}(\lambda)=\begin{bmatrix}[W_{1}(\lambda)]_{+}\\{}
[W_{1}(\lambda)]_{-}\\
W_{-1}(\lambda)
\end{bmatrix}
\]
 onto a constant, the event referred to in \eqref{Wrkdef} holds only
if (for a given realisation) there exists a $b=(b_{1},b_{-1}^{\trans})^{\trans}\in\reals^{q+2}$
such that
\[
0=b^{\trans}\begin{bmatrix}1\\
W^{\ast}(\lambda)
\end{bmatrix}=b_{1}+b_{-1}^{\trans}W^{\ast}(\lambda)
\]
for all $\lambda\in[0,1]$. Taking $\lambda=0$, we see this implies
$b_{1}=0$. Hence it suffices for \eqref{Wrkdef} to show that
\begin{equation}
\Prob\{\exists a\in\reals^{q+1}\text{ s.t. }a^{\trans}W^{\ast}(\lambda)=0\sep\forall\lambda\in[0,1]\}=0.\label{eq:Wrkdef2}
\end{equation}

To that end, we note that by Tanaka's formula (Theorem~VI.1.2 in
\citealp{RY99book}) that
\[
[W_{1}(\lambda)]_{\pm}=\int_{0}^{\lambda}\indic^{\pm}[W_{1}(s)]\diff W_{1}(s)+\frac{1}{2}L_{W_{1}}(\lambda,0)
\]
where $L_{W_{1}}(\lambda,x)$ denotes the local time of $W_{1}$ at
time $\lambda\in[0,1]$ and spatial point $x\in\reals$, which is
a continuous increasing process (for each $x$ fixed). It follows
that $W^{\ast}$ is a vector semimartingale, with quadratic variation
process
\[
Q(\lambda)\defeq\begin{bmatrix}\int_{0}^{\lambda}\indic^{+}[W_{1}(s)]\diff s & 0 & 0\\
0 & \int_{0}^{\lambda}\indic^{-}[W_{1}(s)]\diff s & 0\\
0 & 0 & \lambda I_{q-1}
\end{bmatrix}.
\]
We note that $Q(1)$ is rank deficient only if one of its first two
diagonal entries are zero, which in turn requires that either $\min_{\lambda\in[0,1]}W_{1}(\lambda)\geq0$
or $\max_{\lambda\in[0,1]}W_{1}(\lambda)\leq0$. But since $W_{1}$
is a standard Brownian motion (initialised at zero), both of these
events have zero probability. It follows by a standard characterisation
of quadratic variation (Definition~IV.1.20 in \citealp{RY99book})
that for $\Delta_{m,i}W^{\ast}\defeq W^{\ast}(\frac{i}{m})-W^{\ast}(\frac{i-1}{m})$
\[
Q_{m}(1)\defeq\sum_{i=1}^{m}\Delta_{m,i}W^{\ast}(\Delta_{m,i}W^{\ast})^{\trans}\inprob Q(1)
\]
as $m\goesto\infty$ and thus, since $W^{\ast}(0)=0$, that
\begin{align*}
 & \Prob\{\exists a\in\reals^{q+1}\text{ s.t. }a^{\trans}W^{\ast}(\lambda)=0\sep\forall\lambda\in[0,1]\}\\
 & \qquad\qquad\qquad\qquad\leq\Prob\{\exists a\in\reals^{q+1}\text{ s.t. }a^{\trans}\Delta_{m,i}W^{\ast}=0\sep\forall i\in\{1,\ldots,m\}\}\\
 & \qquad\qquad\qquad\qquad=\Prob\{\rank Q_{m}(1)<q+1\}\\
 & \qquad\qquad\qquad\qquad\goesto0
\end{align*}
as $m\goesto\infty$. Thus \eqref{Wrkdef2} holds.
\end{proof}

\end{document}